\documentclass[fleqn,usenatbib]{mnras}
\usepackage{newtxtext,newtxmath}
\usepackage[T1]{fontenc}
\usepackage{ae,aecompl}
\usepackage{graphicx}   
\usepackage{amsmath}    
\usepackage{amssymb}    
\usepackage{tabularx}
\hypersetup{final}

\title[C-EAGLE velocity dispersion]{The Cluster-EAGLE project: velocity bias and the velocity dispersion - mass relation of cluster galaxies}

\author[T. J. Armitage. et al.]{Thomas J. Armitage,$^{1}$\thanks{E-mail: thomas.armitage-3@postgrad.manchester.ac.uk}
David. J. Barnes$^{1}$,
Scott. T. Kay$^{1}$,
Yannick M. Bah\'e$^{2}$,
\newauthor
 Claudio Dalla Vecchia$^{3,4}$,
Robert A. Crain$^{5}$ and
Tom Theuns$^{6}$
\\
$^{1}$Jodrell Bank Centre for Astrophysics, School of Physics and Astronomy, The University of Manchester, Manchester M13 9PL, UK\\
$^{2}$Max-Planck-Institut f\"ur Astrophysik, Karl-Schwarzschild Str. 1, 85748 Garching, Germany \\
$^{3}$Instituto de Astrof\'\i{}sica de Canarias, C/ V\'\i{}a L\'actea s/n, 38205 La Laguna, Tenerife, Spain\\
$^{4}$Departamento de Astrof\'\i{}sica, Universidad de La Laguna, Av.~del Astrof\'\i{}sico Francisco S\'anchez s/n, 38206 La Laguna,
Tenerife, Spain\\
$^{5}$Astrophysics Research Institute, Liverpool John Moores University, 146 Brownlow Hill, Liverpool, L3 5RF, UK \\
$^{6}$Institute for Computational Cosmology, Department of Physics, University of Durham, South Road, Durham DH1 3LE, UK
}

\date{Accepted XXX. Received YYY; in original form ZZZ}

\pubyear{2017}

\begin{document}
\label{firstpage}
\pagerange{\pageref{firstpage}--\pageref{lastpage}}
\maketitle

\begin{abstract}
We use the Cluster-EAGLE simulations to explore the velocity bias introduced when using galaxies, rather than dark matter particles, to estimate the velocity dispersion of a galaxy cluster, a property known to be tightly correlated with cluster mass. The simulations consist of 30 clusters spanning a mass range $14.0 \le \log_{10}(M_{\rm 200c}/\mathrm{M_\odot}) \le 15.4$, with their sophisticated sub-grid physics modelling and high numerical resolution (sub-kpc gravitational softening) making them ideal for this purpose. We find that selecting galaxies by their total mass results in a velocity dispersion that is 5-10 per cent higher than the dark matter particles. However, selecting galaxies by their stellar mass results in an almost unbiased ($<5$ per cent) estimator of the velocity dispersion. This result holds out to $z=1.5$ and is relatively insensitive to the choice of cluster aperture, varying by less than 5 per cent between $r_{\rm 500c}$ and $r_{\rm 200m}$. We show that the velocity bias is a function of the time spent by a galaxy inside the cluster environment. Selecting galaxies by their total mass results in a larger bias because a larger fraction of objects have only recently entered the cluster and these have a velocity bias above unity. Galaxies that entered more than $4 \, \mathrm{Gyr}$ ago become progressively colder with time, as expected from dynamical friction. We conclude that velocity bias should not be a major issue when estimating cluster masses from kinematic methods.

\end{abstract}

\begin{keywords}
galaxies: clusters: general - galaxies: kinematics and dynamics - methods: numerical
\end{keywords}

\section{Introduction}
Galaxy clusters form from the largest primordial density perturbations to have collapsed by the current epoch. They have great potential to be powerful cosmological probes, as they trace the high mass tail of the halo mass function and their abundance with mass and redshift is sensitive to  cosmological parameters (see \citealt{Allen2011,Kravtsov2012,Weinberg2013,Mantz2014}). 
In order to extract this cosmological information from clusters it is important to have a reliable method to measure cluster masses. As cluster mass is not directly observable, several techniques have been developed using X-ray observations, gravitational lensing or galaxy kinematics. A drawback to these methods is that they are observationally expensive to perform, requiring high quality datasets, and are susceptible to biases due to the simplifying assumptions that have to be made (e.g. spherical symmetry, hydrostatic equilibrium, galaxies as tracers of the underlying mass distribution). 

An observationally cheaper approach is to use cluster scaling relations, observing properties that scale closely with cluster mass \citep{Kaiser1986}. As such, considerable effort has been put into identifying such observables that scale tightly with cluster mass. Examples of observables commonly used as mass proxies are X-ray luminosity, temperature, and 
$Y_{\rm X}$, the product of X-ray temperature and gas mass, 
(e.g. \citealt{Arnaud2007,Vikhlinin2009,Pratt2009,Mantz2016}), 
the Sunyaev-Zel'dovich (SZ) flux 
(e.g. \citealt{Planck2014XX,Saliwanchik2015}),
optical richness 
(e.g. \citealt{Yee2003,Simet2017}),
and the velocity dispersion, $\sigma$, of member galaxies
(e.g. \citealt{Zhang2011,Bocquet2015,Sereno2015}).
In a comparison of scaling relations calibrated using weak lensing masses, $M_{\mathrm{WL}}$, \cite{Sereno2015} found that the intrinsic scatter in the
$\sigma-M_{\mathrm{WL}}$ relation was $\sim14\%$ as opposed to $\sim 30\%,\sim 25\%,$ and $\sim 40\%$ for X-ray luminosity, SZ flux and optical richness respectively. The low scatter in the $\sigma-M$ relation makes it a prime candidate for obtaining relatively cheap cluster mass estimates. 
This result is corroborated by results from numerical simulations, which find the $\sigma-M$ relation for dark matter (DM) particles to be close to the expected virial scaling ($\sigma \propto M^{1/3}$) with minimal scatter, of order 5 per-cent in DM only simulations, and insensitive to cosmological parameters \citep{Evrard2008}. As the velocity dispersion of DM particles is not observable galaxies are instead used as tracers, it is thus important to establish whether galaxies are a fair tracer of the `true' universal DM relation.

In the near future, deep spectroscopic surveys (e.g. with eBOSS, DESI and Euclid), will yield extremely large datasets of galaxy spectra. For example, Euclid is expected to find $\sim 6\times 10^4$ clusters with $S/N>3$ and will obtain $5\times 10^7$ galaxy spectra \citep{Laureijs2011}. With so many clusters, it will likely be the systematics that dominate the error budget. In this paper we focus on one 
particular systematic, namely the velocity {\it bias} that arises from using satellite galaxies, i.e. galaxies residing in the cluster, as tracers of the DM particles.
Velocity bias, which we define here as $b_v=\sigma_{\rm gal}/\sigma_{\rm DM}$, can arise 
from the inclusion of galaxies that are falling into the cluster for the first time. For a virialised galaxy population there are also two main effects 
that act to bias the galaxies relative to the DM. The first mechanism is tidal stripping, where the tidal forces distort and stretch the satellite, 
causing mass loss. The more extended DM (subhalo) component is more susceptible to tidal stripping than the galaxy as it is less bound. The tidal stripping rate depends on the orbital energy of the satellites, so those with low velocities will be preferentially disrupted and removed (\citealt{Ghigna1998,Taffoni2003,Kravtsov2004,Diemand2004}). This biases the velocity dispersion of the remaining 
satellites high relative to the DM particles. The second mechanism that plays a role is dynamical friction (\citealt{Chandrasekhar1943,Esquivel2007}), which reduces the orbital velocity of galaxies, particularly the largest and easiest to observe galaxies, and leads to a lower velocity dispersion.

Velocity bias in clusters has been studied extensively using numerical simulations, with the 
velocity dispersion of DM subhaloes in DM-only simulations 
(e.g. \citealt{Carlberg1994,Gao2004,Diemand2004,Faltenbacher2006,Faltenbacher2007,White2010,Guo2015}), 
or with galaxies using semi-analytic models
(e.g. \citealt{Diaferio2001,Springel2001,Old2013,Saro2013})
or using hydrodynamical simulations that model the formation of galaxies directly 
(\citealt{Frenk1996,Faltenbacher2005,Biviano2006,Lau2010,Munari2013,Caldwell2016}). 
Recent works find velocity dispersions of galaxies/subhaloes that are typically within 10 per cent of the DM values and the value of $b_v$ 
appears to depend on both the sample selection and the implementation of baryonic physics \citep{Evrard2008,Lau2010,Munari2013}. 
For example, \cite{Lau2010} and \cite{Munari2013} both found that selecting galaxies based on their stellar mass, rather than total subhalo mass, 
can reduce the bias. \cite{Lau2010} found the bias reduced from $b_v=1.067 \pm 0.021$ for subhaloes to $b_v=1.029 \pm 0.022$ for 
galaxies at $z=0$, in their simulations with cooling and star formation (CSF) 
but no AGN feedback. \cite{Munari2013} found no significant reduction (from $b_v=1.079 \pm 0.006$ to $b_v=1.078 \pm 0.007$) in their 
CSF simulations, but when AGN feedback was included the bias went from $b_v=1.095 \pm 0.006$ to $b_v=1.075 \pm 0.006$ 
(taking the ratio of their best-fit $\sigma$ values at the pivot mass scale, when averaged over 8 redshift bins from $z=0$ to $z=2$). 
Selecting subhaloes by their mass at infall also produces a similar effect as the stellar and infall mass are more closely related as the DM particles in the subhaloes are more likely to be stripped \citep{Wetzel2010}.

In this paper, we study the velocity bias of galaxies and subhaloes using a new set of cluster simulations, known as the 
Cluster-EAGLE (C-EAGLE) project \citep{Barnes2017b,Bahe2017}. The simulations improve on previous work in a number of ways. Firstly, the resolution of the {C-EAGLE} clusters is significantly higher than in previous work using hydrodynamical simulations studying the velocity dispersion of clusters, with a gas particle mass of $\sim 2 \times 10^6 \, \mathrm{M_\odot}$ as opposed to $2 \times 10^8 \, \mathrm{M_\odot}$ in \cite{Munari2013} and $1 \times 10^9 \, \mathrm{M_\odot}$ in \cite{Caldwell2016}, for example. Secondly, the simulations used 
the EAGLE subgrid physics model which is calibrated to yield realistic field galaxies, based on the 
stellar mass function and size-mass relation \citep{Schaye2015,Crain2015}. This model has also been shown to produce realistic galaxies in the dense cluster environment \citep{Bahe2017} with a broadly realistic intra-cluster medium (ICM) beyond the cluster core \citep{Barnes2017b}. As the galaxies are resolved on kpc scales, down to stellar masses of $\sim 10^9 \, \mathrm{M_\odot}$, processes such as tidal stripping should be more accurately modelled, due to the majority of stars being correctly located deep in the subhalo potential. Having a realistic population of galaxies allows us to identify the intrinsic biases of the $\sigma - M$ relation and which observables are more reliable. We explore the reliability of the tracer galaxies as a function of several properties such as their total mass, stellar mass and time spent inside the cluster. 


The rest of the paper is organised as follows; Section \ref{sims} describes the C-EAGLE simulations in more detail and outlines our method for extracting the galaxies and measuring $\sigma$. In Section \ref{Results} we present our results, showing how the velocity bias depends on the selection criteria and examine its likely origin by studying the time since infall. Finally, in Section \ref{discuss} we summarise and discuss our findings.

\section{C-EAGLE simulations} \label{sims}
Here we provide a brief overview of the C-EAGLE simulations, as well as details of the auxiliary datasets used in this paper. We also briefly describe the sub-grid physics model used in the simulations. For a more comprehensive overview see \cite{Barnes2017b} and \cite{Bahe2017}.

\subsection{Cluster sample and initial conditions}
The C-EAGLE project is comprised of 30 cluster zooms, labelled CE-00 to CE-29, selected from a large ($3.2 \, \mathrm{Gpc}$) parent simulation, details of which can be found in \cite{Barnes2017}. A box of such volume ensures that the most massive and rarest objects expected to form within the observable horizon of a $\Lambda$CDM cosmology are captured, giving a sizeable population of massive galaxy clusters. In total 185,150 haloes with $M_{\rm 200c}>10^{14} \, \mathrm{M}_\odot$\footnote{We define $M_{\rm 200c}$ as the mass enclosed within a sphere of radius $r_{\rm 200c}$ whose mean density is $200$ times the critical density of the Universe.}
were found at $z=0$. These haloes were then binned into 10 evenly-spaced $\log_{10}$ mass bins between 
$14.0 \le \log_{10} (M_{\rm 200c}/\mathrm{M_\odot}) \le 15.4$. This procedure ensures the sample was not biased towards low-mass haloes due to the steep mass function. Haloes with a more massive companion within the maximum of $30 \, \mathrm{Mpc}$ or $20 \, r_{\rm 200c}$ were discarded and 3 haloes from each bin were randomly selected, ensuring a representative sample across the cluster mass range.

The 30 clusters were then re-simulated with DM and baryons using the zoom technique \citep{Katz1993,Tormen1997}, at the reference  EAGLE resolution \citep{Schaye2015} with a DM particle mass $m_\mathrm{DM} \approx 9.7 \times 10^6 \, \mathrm{M_\odot}$ and an initial gas particle mass $m_\mathrm{gas}=1.8 \times 10^6 \, \mathrm{M_\odot}$, we refer to this set of simulations as C-EAGLE-GAS\footnote{A dark matter only (DMO) version of each cluster was also run to quantify the effect of introducing baryons. For these runs, the particle mass was set to $1.15\times 10^{7} \, \mathrm{M}_\odot$. We refer to these clusters collectively as C-EAGLE-DMO.}.
The gravitational softening length was set to $2.66$ co-moving $\mathrm{kpc}$ until $z=2.8$ and $0.70$ physical $\mathrm{kpc}$ at lower redshift. We assumed a flat $\Lambda\mathrm{CDM}$ cosmology based on the \textit{Planck} 2013 results combined with baryonic acoustic oscillations, WMAP polarization and high multipole moments experiments \citep{Planck2014I}. The cosmological parameters are $\Omega_{\rm{b}}=0.04825$, $\Omega_{\rm{m}}=0.307$, $\Omega_{\Lambda}=0.693$, $h\equiv H_0/(100\,\rm{km}\,\rm{s}^{-1}\,\rm{Mpc}^{-1})=0.6777$, $\sigma_{8}=0.8288$, $n_{\rm{s}}=0.9611$ and $Y=0.248$. The initial size of each high resolution region was defined 
such that no low resolution particles fell within $5 r_{\rm 200c}$ of the centre of each cluster at $z=0$,
where the centre is defined as the particle with the most negative gravitational potential. In the \textit{Hydrangea} sample the extent of the high-resolution volume 
is extended further, with no low-resolution particles within $10r_{\rm 200c}$ at $z=0$ 
for 24 of the 30 clusters \citep{Bahe2017}. We use 13 of these clusters in this work (the other 11 were already run with the smaller region) but do not make use of their extra volume as we mainly focus on the volume inside $r_{\rm 200c}$. 
The basic properties of the clusters relevant to this study are listed in Table~\ref{Table:ClusterProps}; see \cite{Barnes2017b} and \cite{Bahe2017} for additional properties.

\subsection{The EAGLE model}

The C-EAGLE simulations were performed using the same model as the EAGLE simulations \citep{Schaye2015,Crain2015}. 
This code is a modified version of the \textit{N}-Body Tree-PM SPH code \textsc{P-Gadget-3}, last described in \cite{Springel2005}. The implemented hydrodynamics is collectively known as \textsc{anarchy} (for details see Appendix A of \citealt{Schaye2015} and \citealt{Schaller2015}). \textsc{anarchy} is based on the pressure-entropy formalism derived by \cite{Hopkins2013} with an artificial viscosity switch \citep{CullenDehnen2010} and includes 
artificial conductivity similar to that suggested by \cite{Price2008}. The $C^2$ smoothing kernel of \cite{Wendland1995} and the time-step limiter of \cite{DurierDallaVecchia2012} are also used.

The EAGLE code is based on that originally developed for the OWLS project \citep{Schaye2010}, also used in the GIMIC \citep{Crain2009}, COSMO-OWLS \citep{LeBrun2014} and BAHAMAS \citep{McCarthy2017,Barnes2017} simulations. This includes radiative cooling, star formation, stellar feedback and the seeding, growth and feedback of black holes. An important advance made for EAGLE was to calibrate the star formation and feedback prescriptions to a limited set of observational data, as these processes cannot be resolved by the simulations properly. The EAGLE model was also calibrated to reproduce the observed relationship between stellar mass and halo mass as well as the size of field galaxies. We briefly summarise the details of each of these key components:

\begin{itemize}
\item Radiative cooling and photo-heating of gas is calculated on an element-by-element basis following \cite{WiersmaSchayeSmith2009} assuming an optically-thin UV/X-ray background along with the cosmic microwave background \citep{HaardtMadau2001}. 

\item Star formation rates are calibrated to reproduce the observed relation with gas surface density in \cite{Kennicutt1998}. This is done using a pressure law \citep{SchayeDallaVecchia2008}, with no star formation below a metallicity-dependent density threshold \citep{Schaye2004}. Each star particle is treated as a simple stellar population with a \cite{Chabrier2003} initial mass function.

\item Feedback from star formation is implemented by injecting thermal energy to the surrounding gas particles of newly created star particles, as described in \cite{DallaVecchia2012}. The energy given to each gas particle is in the form of a fixed temperature change $\Delta T_\mathrm{SF}=10^{7.5} \, \mathrm{K}$. 
The efficiency of the stellar feedback is a function of density and metallicity. The former is designed to further mitigate artificially high radiative losses that are particularly severe in the densest gas where the cooling times are shortest \citep{Schaye2015}. The latter is physically motivated as the cooling time varies inversely with the gas metallicity.
These parameters are calibrated to reproduce the observed galaxy stellar mass function and galaxy size-mass relation at $z=0.1$ \citep{Crain2015}.

\item The seeding, growth and feedback from black holes is based on the method introduced by \cite{Springel2005}, incorporating the modifications of \cite{BoothSchaye2009} and \cite{RosasGuevara2015} accounting for the conservation of angular momentum from accretion. Each DM halo with a mass $>10^{10} \, \mathrm{M_\odot}/h$ has its most bound particle converted into a black hole seed  of mass $10^{5} \, \mathrm{M_\odot}/h$. The black holes can grow either by merging with other black holes, or by accretion of gas at the minimum of the Bondi-Hoyle and Eddington rates \citep{Bondi1944}. Feedback from accreting black holes is modelled in a similar way to the stellar feedback, with the efficiency calibrated to reproduce the locally-observed relationship between the stellar mass of galaxies and the mass of their central super-massive black hole. \cite{Schaye2015} presented three calibrated models for the EAGLE simulations; for C-EAGLE the AGNdT9 model, with 
a heating temperature $\Delta T_\mathrm{AGN}=10^{9} \, \mathrm{K}$, was used. As shown in \cite{Schaye2015}, this model provides a better match to 
the observed gas fractions and X-ray luminosities of low redshift groups than the reference EAGLE model. 
\end{itemize}

The C-EAGLE simulations broadly match many observed properties of the intra-cluster medium (ICM) such as X-ray temperature, luminosity and metallicity \citep{Barnes2017b,Bahe2017}. However, unlike the lower-mass groups, the hot gas fractions in the cluster are too high by $\approx 30$ per cent. Furthermore, the entropy of the gas is also too high in the cluster cores \citep{Barnes2017b}. The stellar masses and the passive fractions of the satellite galaxy population are generally consistent with observations, although the brightest central galaxies (BCGs) are too massive by a factor of $\sim 3$ \citep{Bahe2017}. However, as we do not include the BCGs in this study, and the cluster potential is dominated by the DM throughout the bulk of the cluster volume, our results should be reasonably robust. 

\subsection{Velocity dispersion calculation} \label{velDispTheory}

In this paper, we calculate the velocity dispersion of both the individual DM particles inside a cluster and the galaxies, which are selected and binned by several different criteria. Galaxies are identified using the SUBFIND algorithm \citep{Springel2001,Dolag2009}, run on all 30 snapshots\footnote{The snapshots were spaced $500 \, \mathrm{Myr}$ apart, with two additional outputs at $z=0.1$ and $z=0.37$ for comparison with the original EAGLE simulations.}
for each cluster. To trace their evolution, and determine when they fall into the cluster, halo merger trees were also created using the method described in \cite{Helly2003}. 

Each galaxy typically resides inside a subhalo which has a net velocity and total mass made up of all the particles (DM, gas, stars, and black holes) bound to it. We bin the galaxies by both their total mass, $M_\mathrm{Sub}$, and stellar mass, $M_*$, with a lower limit of $10^9 \, \mathrm{M_\odot}$, as the EAGLE model was not calibrated to reproduce the galaxy population below this limit. This mass threshold ensures all galaxies contain at least 100 particles.

We calculate the velocity dispersion using the gapper method, which was found by \cite{Beers1990} to be robust down to as few as 5 members, an important consideration when calculating $\sigma$ in clusters with few high-mass galaxies
\footnote{We have also performed the analysis using the bi-weight estimator, as suggested by \cite{Beers1990} for large datasets. We found little change in the results and so only present the results from the gapper method in this paper.}. Hence we require at least 5 galaxies to compute the dispersion. For this method, the velocities, $v$, of the $N$ member objects (particles or galaxies) 
are first sorted in increasing size. The velocity dispersion is then calculated using
\begin{equation}
\sigma_{\mathrm{gap}}=\frac{\sqrt{\pi}}{N(N-1)} \sum_{i=1}^{N-1} i(N-i)(v_{i+1} - v_i).
\label{sGapper}
\end{equation}
When calculating the velocity dispersion of galaxies, the central brightest cluster galaxy (BCG) was omitted as it does not orbit in the cluster potential, making it a special case, and unless otherwise stated we calculate the velocity dispersion using the tracers 3D velocity, combining the components together using $\sigma = \sqrt{(\sigma_x^2+\sigma_y^2+\sigma_z^2)/3}$. We also use the true radial positions of galaxies when making radial cuts, as we do not consider projection effects or the impact of interlopers in this work.

\section{Results} \label{Results}

\subsection{The effect of baryons on the DM velocity dispersion}

We first assess the effect of baryons by comparing the velocity dispersions of the DM particles in the 
C-EAGLE-GAS and C-EAGLE-DMO simulations. To do this, we parametrise the $\sigma-M$ relation following \cite{Evrard2008}
\begin{equation}
    \log_{10} (\sigma_{\rm 200c})= \log_{10} ( \sigma_{\mathrm{piv}} ) + \alpha \log_{10} \bigg(\frac{h(z)M_{\rm 200c}}{M_{\mathrm{piv}}} \bigg),
    \label{s200M200}
\end{equation}
where $h(z)=H(z)/100 \, \mathrm{km/s}$ is the dimensionless Hubble parameter.
We set the pivot mass to $M_{\mathrm{piv}}=4\times10^{14} \, \mathrm{M}_{\odot}$, chosen to minimise the covariance between the normalisation, 
$\sigma_{\mathrm{piv}}$, and gradient, $\alpha$, parameters. Note that for a self-similar cluster population, $\alpha$ would be $1/3$, and both 
$\sigma_\mathrm{piv}$ and $\alpha$ would be independent of redshift. 
We also quantify the amount of scatter, $\delta$, in the $\sigma - M$ relation using
\begin{equation}
    \delta =\sqrt{\frac{1}{N} \sum_{i=1}^{N} \log_{10}(\sigma_i / \sigma_\mathrm{fit})^2 },
\label{Scatter}
\end{equation}
where $\sigma_i$ is the velocity dispersion of the $i^{\rm th}$ cluster and $\sigma_\mathrm{fit}$ is the best-fit value given its $M_{\rm 200c}$. 
The results of these fits are presented in Table \ref{z0Params}, where we present the scatter in the form of $\delta_{\ln} = \ln(10)\delta$ for ease of
comparison with previous work. The statistical uncertainty on each parameter is estimated using the bootstrap method, based on 10,000 re-samples of the data.  
Additional results for other limiting radii are considered in Section \ref{aperture_var}.

\begin{figure}
    \centering
    \includegraphics[width=1.0\linewidth]{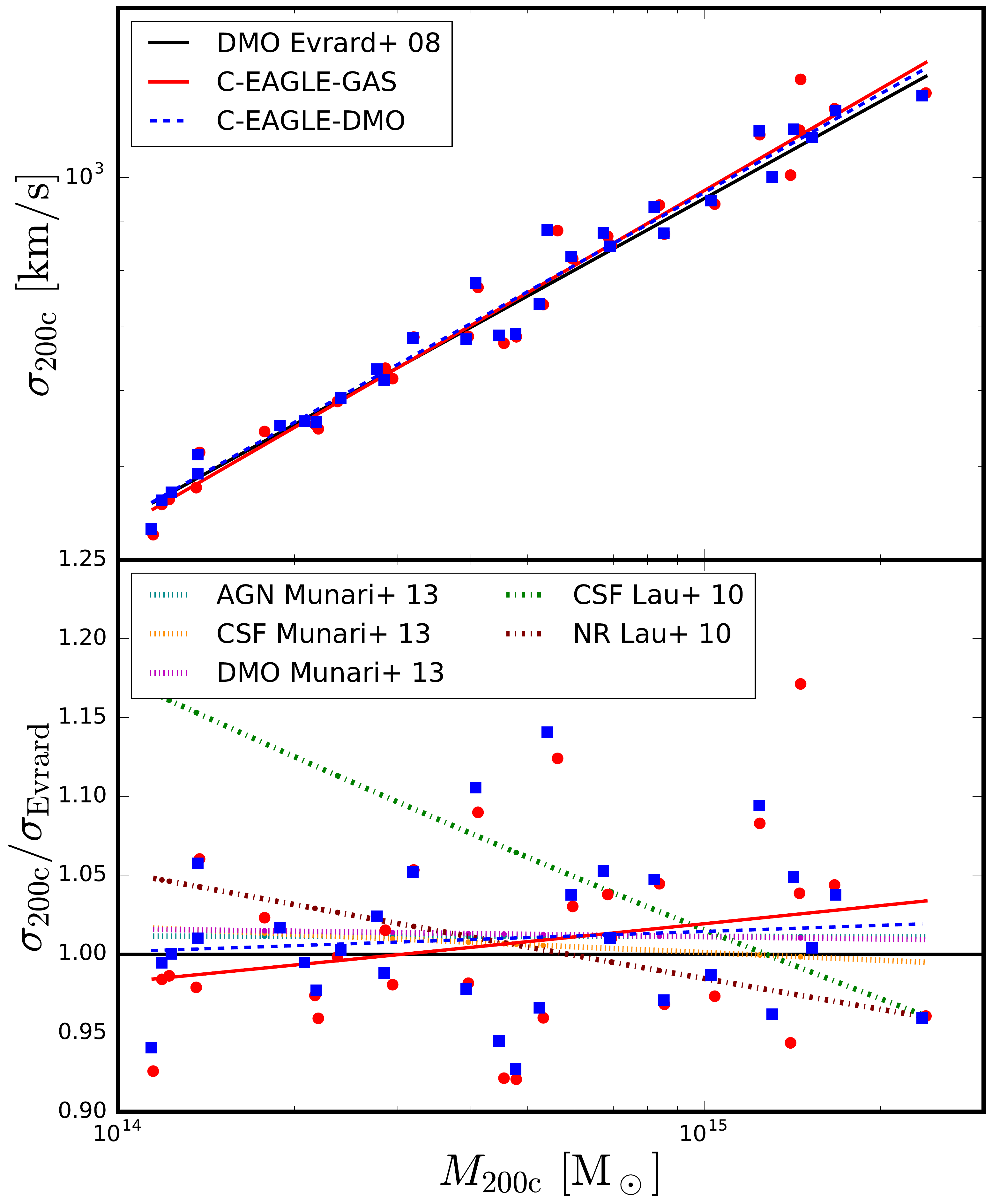}
    \caption{Top panel: the $\sigma_{\rm 200c} - M_{\rm 200c}$ relation for all 30 C-EAGLE clusters at $z=0$. The red solid and dashed blue lines show the best-fit relation for the DM particles  
    within $r_{\rm 200c}$ for the C-EAGLE-GAS and C-EAGLE-DMO runs respectively. Similarly, the blue squares and red circles show results for individual clusters in the two simulation sets. 
    The solid black line is the best-fit relation found by \citet{Evrard2008}. Bottom panel: best-fit relations relative to the \citet{Evrard2008} result. Additionally, the dotted magenta, orange and 
    cyan lines are the best-fit results found by \citet{Munari2013} for their DMO, CSF and AGN runs respectively, while the burgundy and green dot-dash lines are for the NR and CSF simulations 
    found by \citet{Lau2010}, respectively.}
    \label{fig:SigmavsmassDMPonly029}
\end{figure}

\begin{table*}
    \centering
    \caption{Best-fit parameters to the $\sigma_{\rm 200c}-M_{\rm 200c}$ relation for the DM particles and galaxies at $z=0$. Column~1 lists the simulation and sample details (with mass limits where appropriate). 
    Column~2 shows the typical number of galaxies in each mass bin for a cluster with $M_\mathrm{200,c}=M_\mathrm{piv}=4\times10^{14} \, \mathrm{M_\odot}.$ 
    Column~3 gives the best-fit slope of the relation and the $1\sigma$ uncertainty. Columns 4-5 give the 
    best-fit normalisation and scatter. Finally, column 6 gives the ratio of the 
    normalisation to the case for DM particles, a measure of the velocity bias for the galaxies.}
    \label{z0Params}
    \begin{tabular}{l c c c c c}
        \hline
         & $N_\mathrm{eff,piv}$ & $\alpha$ & $\sigma_{\mathrm{piv}} \, [\mathrm{km/s}]$ &  $\delta_{\ln} $ & $\sigma_\mathrm{piv}/\sigma_\mathrm{piv,DM}$ \\ 

         \hline
    
    C-EAGLE-DMO\\
         DM Particles & & $0.34 \pm 0.01$ & $805 \pm 8 $ & $0.048 \pm 0.006$\\ 

         $M_\mathrm{Sub}:$ $10^9~-10^{10} \,  \mathrm{M_\odot}$ & $2419 \pm 237$ & $0.32 \pm 0.01$ & $887 \pm 7 $ & $0.035 \pm 0.005$ & $1.10 \pm 0.01$\\ 

         $M_\mathrm{Sub}:$ $10^{10}-10^{11} \, \mathrm{M_\odot}$ & $\phantom{0}286 \pm \phantom{0}36$ & $0.33 \pm 0.01$ & $893 \pm 7 $ & $0.033 \pm 0.004$ & $1.11 \pm 0.01$\\ 

         $M_\mathrm{Sub}:$ $10^{11}-10^{12} \, \mathrm{M_\odot}$ & $\phantom{00}30 \pm \phantom{00}6$ & $0.33 \pm 0.02$ & $\phantom{0}894 \pm 13 $ & $0.062 \pm 0.008$ & $1.11 \pm 0.02$\\

         \hline

    C-EAGLE-GAS\\
         DM Particles & & $0.35 \pm 0.01$ &  $804 \pm 9 $ & $0.055 \pm 0.007$\\ 

         $M_\mathrm{Sub}:$ $10^9~-10^{10} \, \mathrm{M_\odot}$ & $1897 \pm 193$ & $0.32 \pm 0.01$ & $883 \pm 7 $ & $0.033 \pm 0.005$ & $1.10 \pm 0.01$\\ 

         $M_\mathrm{Sub}:$ $10^{10}-10^{11} \, \mathrm{M_\odot}$ & $\phantom{0}225 \pm \phantom{0}23$ &$0.33 \pm 0.01$ & $885 \pm 8 $ & $0.034 \pm 0.004$ & $1.10 \pm 0.02$\\ 

         $M_\mathrm{Sub}:$ $10^{11}-10^{12} \, \mathrm{M_\odot}$ & $\phantom{00}37 \pm \phantom{00}6$ & $\phantom{00}0.34 \pm 0.02$ & $\phantom{0}844 \pm 11 $ & $0.08\phantom{0} \pm 0.01\phantom{0}$ & $1.05 \pm 0.02$\\ 


         $M_{\mathrm{*}}\,\,\,\,\,:$ $10^{9~}-10^{10} \, \mathrm{M_\odot}$ & $\phantom{00}90 \pm \phantom{0}11$ & $0.32 \pm 0.02$ & $\phantom{0}823 \pm 11 $ & $0.06\phantom{0} \pm 0.01\phantom{0}$ & $1.02 \pm 0.02$\\ 

         $M_{\mathrm{*}}\,\,\,\,\,:$ $10^{10}-10^{11} \, \mathrm{M_\odot}$ & $\phantom{00}42 \pm \phantom{00}6$ & $\phantom{00}0.36 \pm 0.02$ & $\phantom{0}789 \pm 13 $ & $0.11\phantom{0} \pm 0.02\phantom{0}$ & $0.98 \pm 0.02$\\

         \hline

    \end{tabular}
\end{table*}

The velocity dispersions of the DM particles inside $r_{\rm 200c}$ are consistent between the C-EAGLE-GAS and C-EAGLE-DMO simulations, as can also be seen in 
Fig. \ref{fig:SigmavsmassDMPonly029}. We also find the relations to be consistent with those of \cite{Munari2013} and \cite{Evrard2008}. In all cases, the slope is 
close to the self-similar value of $\alpha=1/3$, as can be seen in Table \ref{z0Params}. (We also present fits where $\alpha$ is fixed at $1/3$ in Table \ref{z0Params_alphaFixed}). The relations found by \cite{Lau2010} differ notably from our results, particularly when radiative cooling and star formation 
(labelled as CSF) are included. This results in a significantly shallower slope ($\alpha=0.27$) compared to their simulation with non-radiative (NR) gas ($\alpha=0.31$). As discussed by 
\cite{Lau2010}, this effect can be explained as being due to the dissipation of baryons, leading to a larger value of $\sigma$, particularly for lower-mass clusters where cooling is more efficient. A similar model is 
considered by \cite{Munari2013} but they do not see such a significant effect. 

\begin{figure*}
    \centering
    \includegraphics[width=1.0\linewidth]{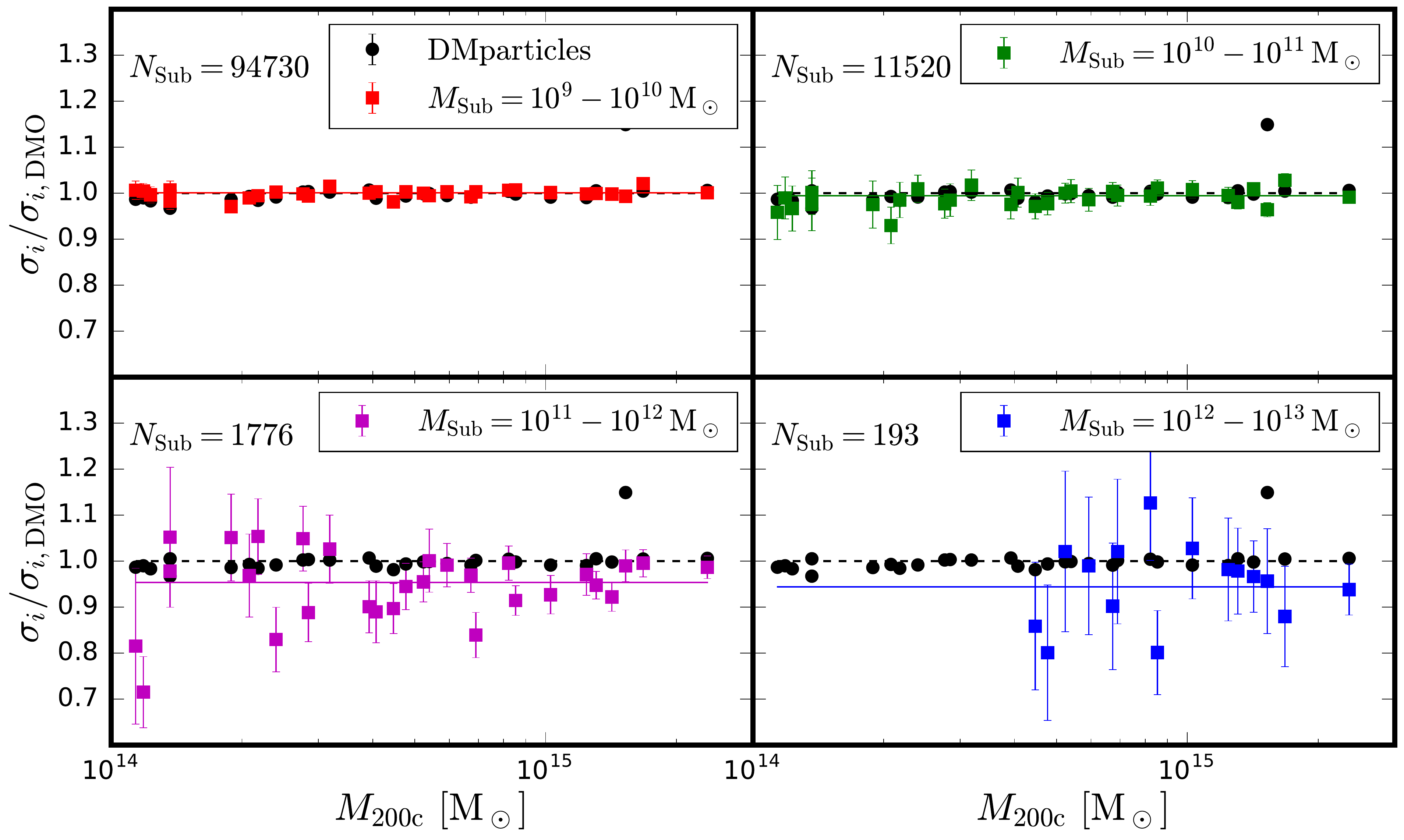}
    \caption{The effect of baryons on the cluster velocity dispersion. Each point shows the ratio of the velocity dispersion for 
    a cluster in the C-EAGLE-GAS to the corresponding object in the C-EAGLE-DMO runs. Black circles are for DM particles within $r_{\rm 200c}$, which are the same in all panels, 
    while coloured squares correspond to subhaloes/galaxies binned by total mass (one bin per panel). Error bars show one standard deviation from 
    bootstrapping the data 10,000 times. The dashed line represents no change between simulations while the solid line gives the
    weighted mean for each subhalo mass bin. The total number of galaxies within the C-EAGLE-GAS clusters in the respective 
    total mass range is also given in each panel. 
    }
    \label{fig:DM_DMO_Comp}
\end{figure*}

We find that the scatter in the $\sigma_{\rm 200c}-M_{\rm 200c}$ relation for the C-EAGLE-DMO clusters, $\delta_{\ln}=0.048 \pm 0.006$, is consistent with 
that found by \cite{Lau2010} for their NR model, $\delta_{\ln}=0.041 \pm 0.008$ and the global relation found by \cite{Evrard2008}, $\delta_{\ln}=0.043 \pm 0.002$. 
We also find that the scatter does not significantly change when baryons are introduced, increasing by one standard deviation from $0.048 \pm 0.006$ to $0.055 \pm 0.007$. This is 
similar to \cite{Lau2010}, who find $\delta_{\ln}=0.056 \pm 0.009$ for their CSF simulation. \cite{Munari2013} consistently found the scatter to be $\sim 0.05$ across their DMO, NR, CSF, and AGN runs. 
These results imply that the scatter is relatively insensitive to any implemented baryonic physics.

In Fig. \ref{fig:DM_DMO_Comp}, we also compare $\sigma_{\rm 200c}$ for individual clusters in the C-EAGLE-GAS and C-EAGLE-DMO simulations. 
The black circles show the ratio, $R$, of the C-EAGLE-GAS and C-EAGLE-DMO velocity dispersions versus $M_{\rm 200c}$ (the same result is shown in each panel). 
The weighted mean ratio is consistent with unity, $\left<  R \right> = 1.00 \pm 0.03$. Note that one cluster (CE-27) is an outlier with $R=1.15$; this object undergoes 
a major merger around $z=0.1$, however due to timing differences in the simulations this does not happen at precisely the same time in the DMO run, changing the times at which SUBFIND considers the two merging clusters the same. At $z=0.1$ the mass ratio is $0.55$, increasing to $0.96$ at $z=0$. Clearly CE-27 is significantly disturbed at low redshift and so it is not surprising that there is still a discrepancy in the DM velocity dispersion at $z=0$. 

\subsection{Velocity bias in galaxies}

\begin{figure*}
    \centering
    \includegraphics[width=0.99 \textwidth]{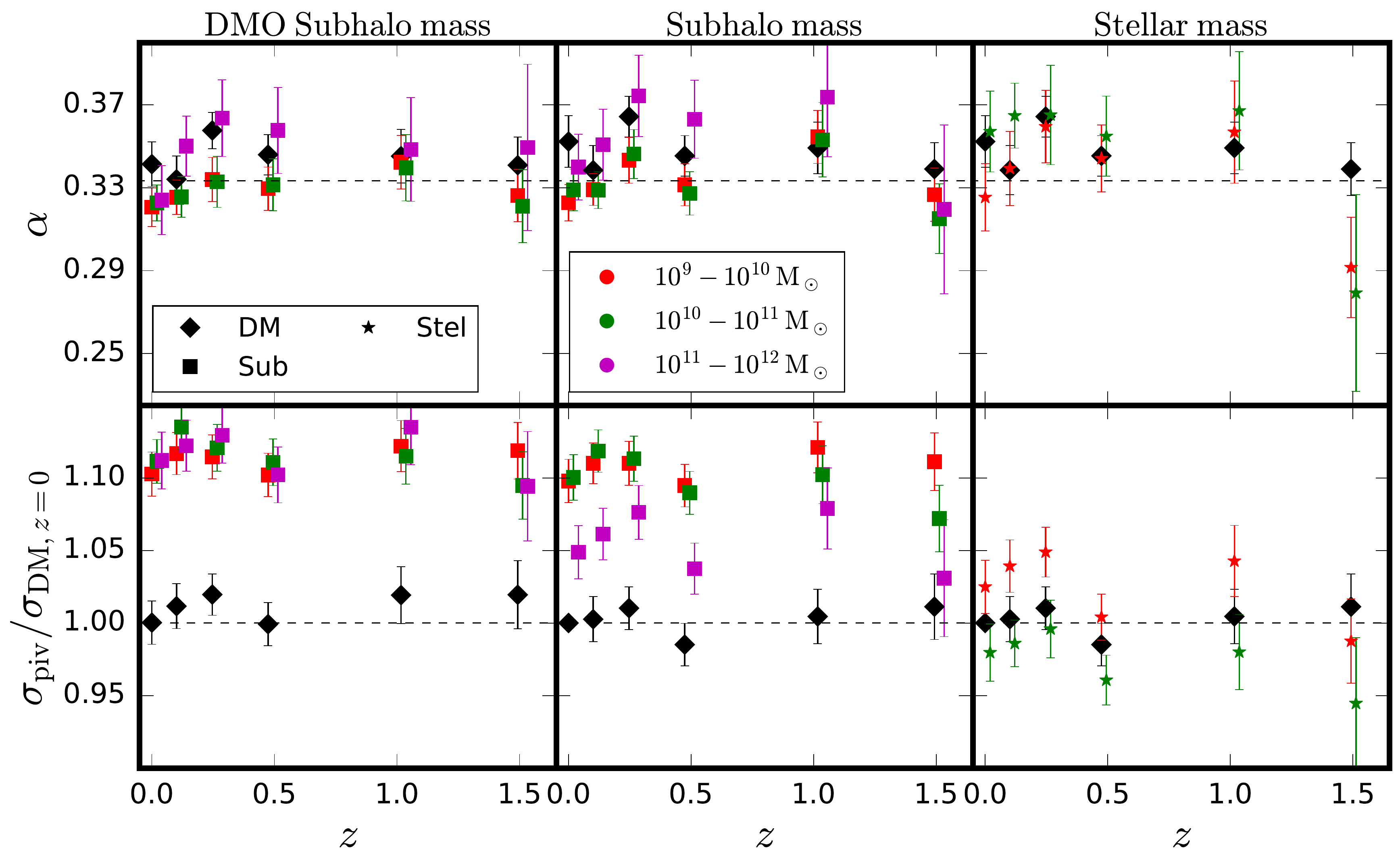}
    \caption{Fitting parameters obtained for Eq. (\ref{s200M200}) using galaxies and DM particles. The top row shows the fitted value of logarithmic slope, $\alpha$, as a function of redshift, while the bottom row shows the values of the normalisation $\sigma_{\mathrm{piv}}$. $\sigma_{\mathrm{piv}}$ is shown relative to the value obtained for the DM particles at $z=0$. The expected self-similar value of $\alpha=1/3$ is shown by the dashed line in the top row, where as the dashed lines in the lower panel show now change between the value of $\sigma_{\mathrm{piv}}$ obtained at $z=0$ using DM particles. The columns left to right are for the DM only runs, with subhaloes binned by total mass, galaxies binned by total mass in the hydro runs and galaxies binned by stellar mass. The uncertainties are obtained by bootstrap re-sampling the 30 clusters $10^4$ times.} 
    \label{fig:ParameterComp}
\end{figure*}

As the velocity dispersion of the DM is not directly observable, it is important to determine if the galaxies residing within the cluster are reliable tracers of the underlying matter distribution. In order to establish how biases come into play, the galaxies are split into several bins depending on their total mass, $M_{\rm Sub}$, and their stellar mass, $M_*$. We also use the term subhalo to refer to objects in the C-EAGLE-DMO sample and when comparing between the GAS and DMO sample we use subhalo to refer to galaxies selected by total mass.

We first look at how the dynamics of the subhaloes change when baryons are introduced. Each panel in Fig. \ref{fig:DM_DMO_Comp} shows the ratio of the subhalo velocity dispersion 
in a cluster between the C-EAGLE-GAS and C-EAGLE-DMO runs, for 4 subhalo mass ranges. (We do not match individual subhaloes between the two simulations of each cluster as this approach is complicated by timing offsets.) 
We can see from the figure that the introduction of baryonic physics has a mass dependent effect on the subhaloes. For subhaloes with masses between $10^9 - 10^{10} \, \mathrm{M_\odot}$ we find no statistically significant change in $\sigma_{\rm 200c}$. Similarly, for $M_{\rm Sub}=10^{10} - 10^{11} \, \mathrm{M_\odot}$, the average change is $\left< R \right> = 0.99 \pm 0.012$. However, for the higher mass subhaloes, which are in the mass range studied in previous work, there is a decrease in $\sigma_{\rm 200c}$ of around 5 per cent, and an increase in the cluster-to-cluster scatter with the introduction of baryons. The decrease in $\sigma$ is also radially dependent, with subhaloes within the inner region ($<0.2 r_{\rm 200c}$) having a lower velocity dispersion in the runs with baryons. Since we know that the central galaxies are too massive in our simulations \citep{Bahe2017} we refrain from drawing conclusions about the central region of the clusters.

Fig. \ref{fig:ParameterComp} shows the best fit parameter values for the $\sigma_{\rm 200c}-M_{\rm 200c}$ relation as a function of redshift, for the different subhalo and galaxy mass bins as well as for the DM particles (the $z=0$ results for the subhaloes and galaxies are also listed in Table~\ref{z0Params}). 

The left and middle columns show the best fit parameters for the total mass bins in the C-EAGLE-DMO and C-EAGLE-GAS runs respectively, whereas the right column shows the best fit parameters for the C-EAGLE-GAS galaxies. The logarithmic slope, $\alpha$, is shown in the top panels, with the dashed horizontal line showing the expected self-similar value of 
$\alpha=1/3$. The lower row shows the normalisation, $\sigma_{\mathrm{piv}}$, relative to the value found for the DM particles in the C-EAGLE-GAS runs at $z=0$. 

In all cases, the slopes are broadly consistent with the expected self-similar value, although there is a tendency for them to be high, particularly the highest mass subhaloes and galaxies at intermediate redshifts ($0.25<z<1$). However, these deviations do not affect the trends seen in the normalisation, as can be deduced from comparing the results in the bottom row of Fig. \ref{fig:ParameterComp}. Our slopes are consistent with those found by \cite{Munari2013} and \cite{Caldwell2016} who obtained slopes of $0.364 \pm 0.0021$ and $0.385 \pm 0.003$ respectively for galaxies with stellar masses $\gtrsim 10^{10} \, \mathrm{M_\odot}$.

When $\sigma_{\rm 200c}$ is calculated using the DM particles, no systematic deviation from self-similar evolution with redshift is seen for the runs with and without baryons for $z<1.5$. This result is consistent with the findings of \cite{Munari2013} and \cite{Caldwell2016} who also find no evolution outside of that expected from self-similar. All subhalo mass bins show a velocity bias of $\sim1.1$ for the C-EAGLE-DMO simulation that persists to $z=1.5$. However, when baryons are included we find the velocity bias decreases for increasing total mass. The smallest mass bins ($M_{\rm Sub}<10^{11} \, \mathrm{M}_{\odot}$) have a similar bias to those in C-EAGLE-DMO but the bias reduces to $\sim 1.05$ for the highest mass bin. This trend with total mass persists to higher redshift.

When selecting galaxies by stellar mass the bias is significantly reduced; our results are consistent with no bias at $z=0$ for both stellar mass bins considered.
There is still a mass trend, with the least massive galaxies having a higher velocity dispersion, and the fitting error is marginally greater relative to the total mass value. Nevertheless, the results obtained suggest that binning the data by stellar mass is a more effective way to reduce the velocity bias between the galaxies and DM particles. This result is in agreement with previous work. As discussed in the introduction, 
\citealt{Munari2013} found a velocity bias of $b_v=1.095$ when using subhaloes 
with mass $>10^{11} \, \mathrm{M_\odot}$ in their full AGN run. When selecting galaxies with stellar mass $>3\times 10^9 \, \mathrm{M_\odot}$ they obtain a smaller 
bias of $b_v=1.075$. While both of these results are higher than what we find for a similar mass range, the trend is the same. A similar result was also found by \citet{Lau2010} for their CSF simulation. 

An unexpected result, seen in Table~\ref{z0Params}, is that the scatter in the $\sigma_{\rm 200c}-M_{\rm 200c}$ relation is lower ($\delta_{\ln}\simeq 0.03$)
for the $10^{9}-10^{11} \, \mathrm{M_\odot}$ galaxies than for the DM particles ($\delta_{\ln}\simeq 0.05$). 
This is the case for both the C-EAGLE-GAS and C-EAGLE-DMO runs and does not change when we require that $\alpha=1/3$. \cite{Munari2013} found that their subhaloes (with $M_{\rm Sub}>10^{11} \, \mathrm{M_\odot}$) had a scatter around twice that of the DM particles. We do not find such a large increase in scatter in the $10^{11}-10^{12} \, \mathrm{M_\odot}$ mass bin, with an increase in scatter of $45 \%$.
Compared to other work \cite{Caldwell2016} find a total scatter of $\delta_{\ln} = 0.16$ in the high cluster mass regions of their simulation sample, which is higher than our value of $\delta_{\ln} = 0.11$ for a similar mass range. \cite{Munari2013} attempted to separate the intrinsic and statistical components due to scatter and concluded that the subhaloes' intrinsic scatter was comparable to the DM particles, i.e. if there were significantly more high mass subhaloes per cluster the scatter would tend to the DM value of $\sim 5$ per cent. Our findings take this further as the galaxies binned by total mass suffer from significantly less scatter than the DM particles, whereas binning in stellar mass results in greater scatter.

The scatter observed in the stellar mass bins, as well as in the $10^{11}-10^{12} \, \mathrm{M_\odot}$ total mass bin, is likely affected by statistical noise as there are significantly smaller number of galaxies present in each cluster for those mass ranges. The second column of Table \ref{z0Params} shows the typical number of galaxies, $N_\mathrm{eff,piv}$, in each bin for a cluster of mass $M_{\rm 200c}=M_\mathrm{piv}=4\times 10^{14} \, \mathrm{M_\odot}$. We calculate this value by dividing the number of galaxies in each mass bin by the host cluster mass, multiplying by the pivot mass and then taking the median and 68 percentile range. We can see that the number of galaxies in each increasing mass bin decreases by an order of magnitude and so it is not surprising that the scatter is larger for the higher mass bins.

Returning to Fig. \ref{fig:DM_DMO_Comp}, we noted that CE-27, which undergoes a major merger at $z=0.1$, has a very different DM velocity dispersion between the DMO and GAS simulations. However, the subhalo velocity dispersion has changed far less.
We follow up on this observation by splitting the clusters into relaxed and unrelaxed objects, using the relaxation criterion in \cite{Barnes2017b}: relaxed clusters are defined as those with $E_\mathrm{kin,500c}/E_\mathrm{therm,500c}<0.1$, where $E_\mathrm{kin,500c}$ is the sum of the kinetic energy of the gas particles inside $r_{\rm 500c}$, excluding any bulk motion, and $E_\mathrm{therm,500c}$ the sum of the thermal energy of the gas particles. Using this criterion, we find that 9 clusters are relaxed at $z=0$. Calculating the scatter for the relaxed sample, we find that for the DM particles it decreases from $\delta_{\ln}=0.055 \pm 0.007$ to $\delta_{\ln}=0.018 \pm 0.003$. The scatter in the subhalo and galaxy relations are less affected by selecting only the relaxed clusters, all bins changed by less than one standard deviation when restricting the sample to relaxed clusters. The exception was the $10^9 - 10^{10} \, \mathrm{M_\odot}$ subhalo bin, which changed from $\delta_{\ln} =0.033 \pm 0.005$ to $\delta_{\ln} =0.019 \pm 0.004$. We conclude that the DM particle velocity dispersion is more susceptible to the past history of the cluster than the subhaloes or galaxies. However, the limited number of galaxies is likely to dominate the measured scatter, rather than the intrinsic variability.

\subsection{Stellar mass - total mass relation}
 \begin{figure*}
    \centering
    \includegraphics[width=0.9 \linewidth]{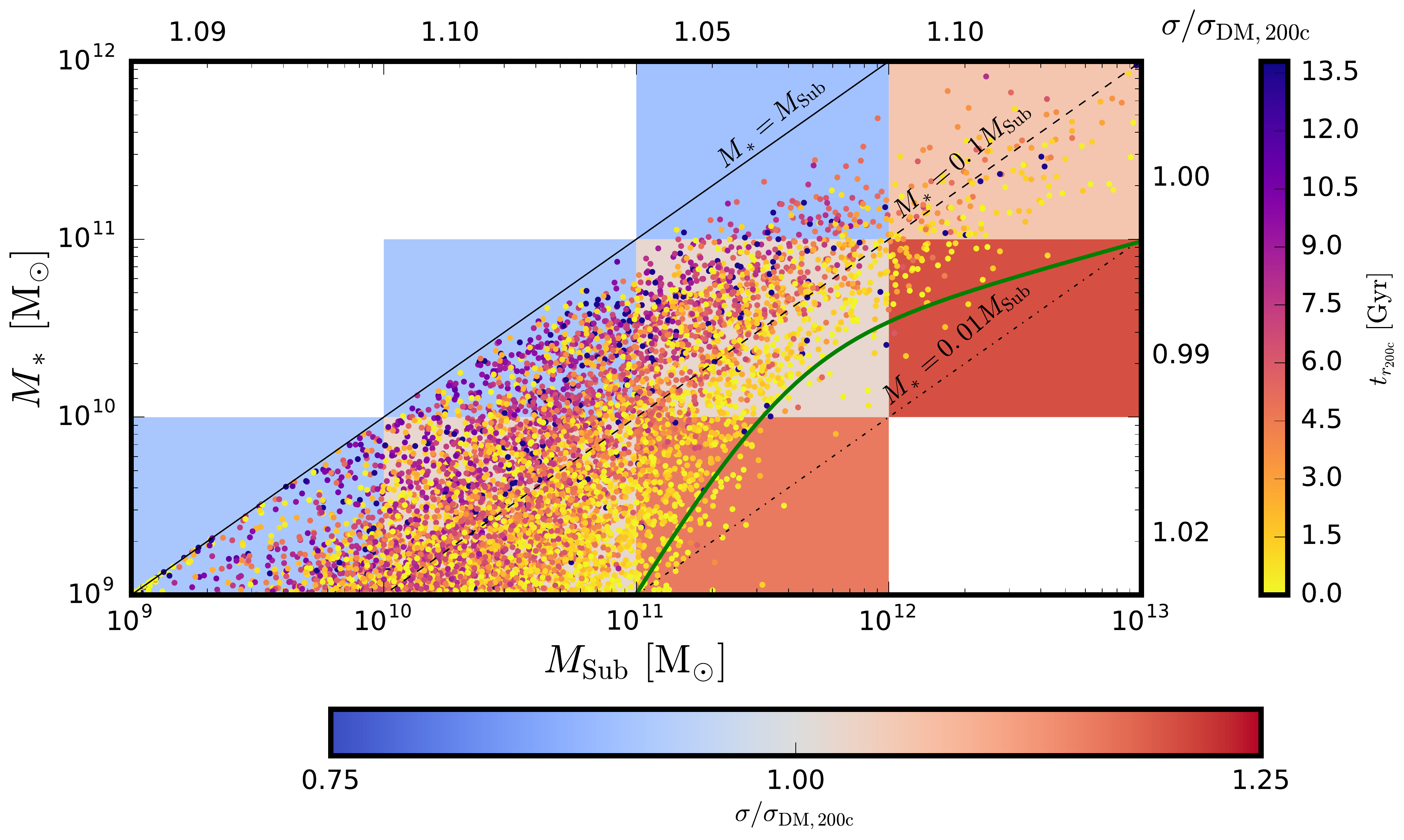}
    \caption{Stellar mass versus total mass for galaxies within $r_{\rm 200c}$ from all 30 clusters. The colour of each point denotes the time since the galaxy first crossed $r_{\rm 200c}$, where a galaxy with $t_\mathrm{r_{200c}=}0$ has only just entered the cluster. The colour grid behind the points illustrates the velocity bias of the galaxies in each 2D bin, while values on the top (right) give the bias for all objects within each total (stellar) mass bin (errors are typically $\pm 0.01$). The black solid, dashed and dot-dashed lines show stellar mass fractions of 1, 0.1 and 0.01 respectively. The green solid line shows the stellar mass-halo mass relation from \citet{Moster2013}.} 
    \label{fig:AllSMassSStel}
\end{figure*}

 \begin{figure*}
    \centering
    \includegraphics[width=0.9 \linewidth]{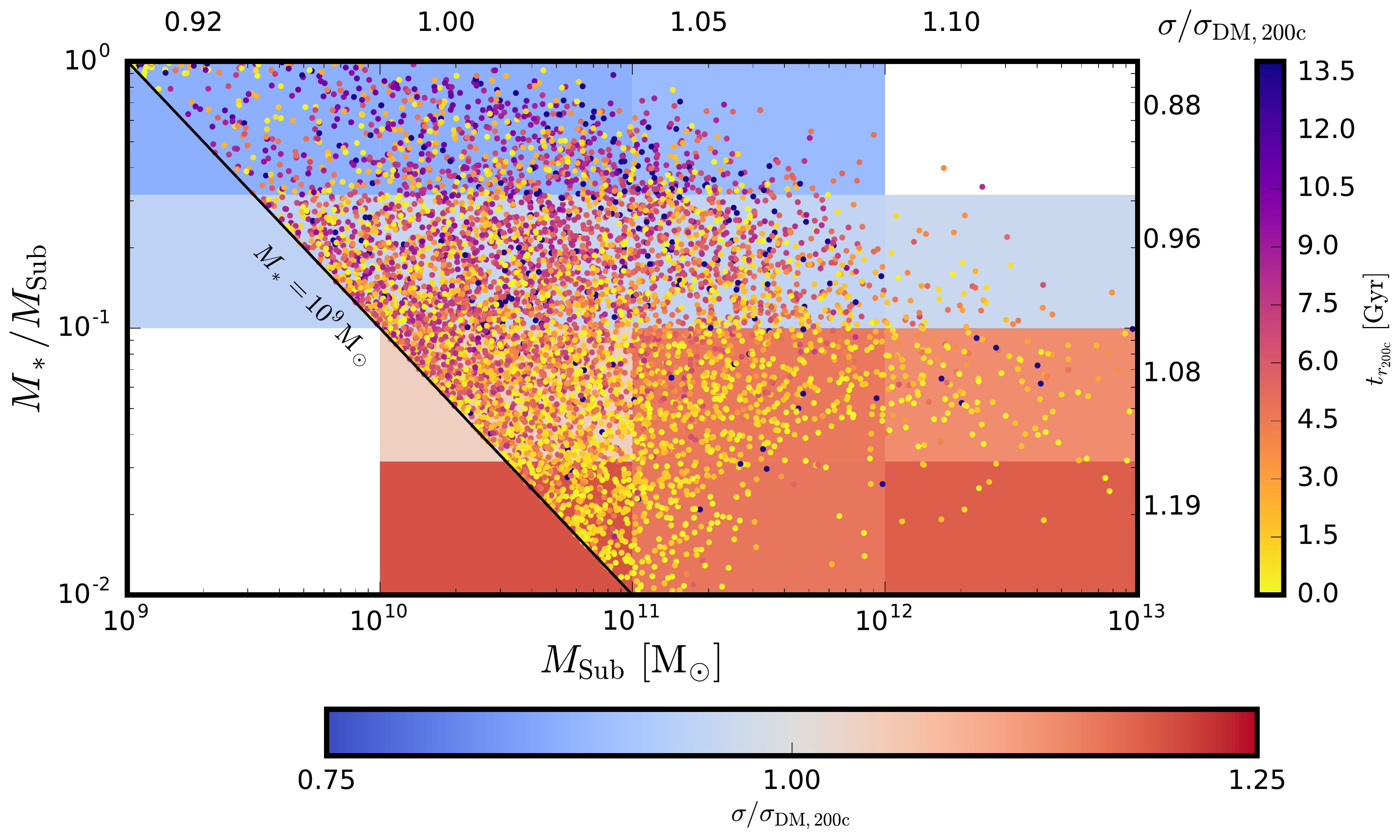}
    \caption{Stellar mass fraction versus total mass for galaxies within $r_{\rm 200c}$ from all 30 clusters. Only galaxies with a stellar mass greater than $10^9 \, \mathrm{M_\odot}$ are shown. The layout is the same as in Fig. \ref{fig:AllSMassSStel}.} 
    \label{fig:AllSMassFrac}
\end{figure*}

We now seek to understand the underlying reason why the velocity bias differs when binning galaxies by their stellar rather than total mass within $r_{\rm 200c}$. Fig.~\ref{fig:AllSMassSStel} shows the stellar mass - total mass relation at $z=0$ for all galaxies in the 30 clusters with both 
$M_\mathrm{Sub}>10^9 \, \mathrm{M_\odot}$ and $M_\mathrm{*}>10^9 \, \mathrm{M_\odot}$. 
The solid, dashed and dot-dashed diagonal lines correspond to stellar mass fractions of $1$, $0.1$ and $0.01$ respectively. The stellar mass-halo mass relation found by \cite{Moster2013} is also shown as the green solid line, which applies to central galaxies and is therefore not directly comparable with a cluster population. We can see that the qualitative shape of the C-EAGLE distribution is similar to that found by \cite{Moster2013}, however it is offset to larger stellar mass fractions. This result is discussed 
in more detail in \cite{Bahe2017}, who found this offset to persist to at least $10r_{\rm 200c}$. 

A significant driver of the scatter in the stellar mass fraction is the time a galaxy has resided within the cluster. We use the merger trees to trace the main progenitors of the galaxies back in time until they first crossed $r_{\rm 200c}(z)$; we define this time as 
$t_{r_{\rm 200c}}$, so that newly entered galaxies have $t_{r_{\rm 200c}} = 0 \, \mathrm{Gyr}$. Older galaxies, i.e. with a larger $t_{r_{\rm 200c}}$, tend to have a higher stellar mass fraction. This is because the  DM is more diffuse than the stellar component and so is more easily stripped due to dynamical processes as the galaxy orbits within the 
cluster potential. 

The velocity bias of all galaxies with respect to the DM particles in their host cluster is also shown in Fig. \ref{fig:AllSMassSStel}. To calculate this, we first scale out the cluster mass dependence by dividing the velocity of 
each galaxy by the velocity dispersion of the DM particles in its host halo. The overall bias, calculated as a 2D 
distribution of both total and stellar mass, is shown on the background grid, with blue (red) regions corresponding to smaller (larger) 
velocity dispersions than the DM particles. We also give values of the bias for each mass bin along the 
top (right) of the figure. For these, all galaxies in the mass bin are taken into account, not just those with masses 
above $10^{9} \, \mathrm{M}_{\odot}$ as shown. This is done to be more consistent with the values in Table \ref{z0Params}.

\begin{figure*}
    \includegraphics[width=0.9 \linewidth]{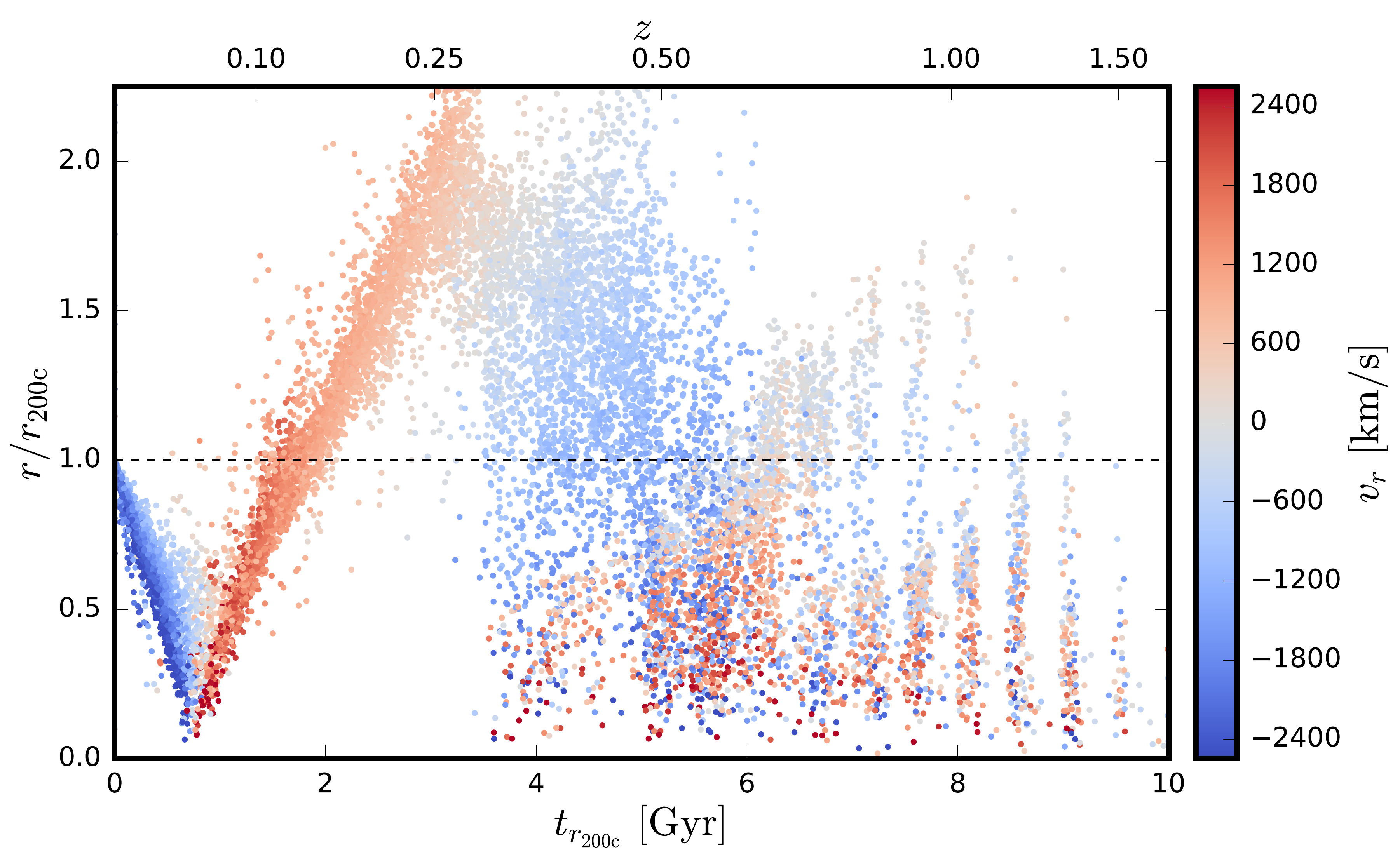}
    \caption{Radial positions of all galaxies at $z=0$ with $M_\mathrm{Sub}>10^9 \, \mathrm{M_\odot}$ in CE-26 against time of first 
    crossing $r_{\rm 200c}$. Colours indicate the radial velocity of each subhalo with respect to the cluster centre and the 
    dashed line is $r_{\rm 200c}$ at $z=0$. The velocity dispersion of the DM particles inside $r_{\rm 200c}$ is 1119 km/s at $z=0$.}
    \label{fig:HM26vrt}
\end{figure*}

There is a clear trend that galaxies that have been inside the cluster longer (and thus have a stellar mass fraction closer to unity) have an accordingly lower velocity dispersion. Fig. \ref{fig:AllSMassSStel} also shows the reduced velocity bias when binning by stellar, rather than total, mass, denoted by the values on the edge of the plot. To understand why the velocity bias is reduced, consider a toy model of a cluster where all galaxies have the same stellar mass fraction upon entering. Galaxies begin to spread out along the $M_{\rm Sub}$ direction only as the stellar mass of a galaxy does not begin to be stripped until a significant fraction ($\sim 80$ per cent) of the surrounding DM has been lost \citep{Smith2016}, due to the stellar component residing much deeper in the subhalo potential. Hence, by selecting objects based on stellar mass one obtains a fairer sample of tracers, with each object at a different point in their dynamical history. Contrast this with slices along the $M_\mathrm{Sub}$ axis, where there will now be a mix of galaxies that are young (with a low stellar mass fraction) and old (with a high stellar mass fraction). Crucially, their dynamical histories (past and future) will be different as this depends on their total mass when they first entered the cluster. Furthermore, due to the steepness of the halo-mass function, results are biased towards newly-entered, dynamically hot galaxies when binning in total mass. This does not occur for stellar mass bins as a subhalo will tend to stay within that bin until close to the end of its life, when the stars eventually become stripped. The key point is that selecting galaxies by total mass results in a younger population in a given mass range compared to selecting by stellar mass. When selecting by stellar mass the older galaxies, which have been slowing down due to dynamical friction happen to compensate for the young hot galaxies, reducing the bias.

We show the correlation between stellar mass fraction, age and velocity bias more clearly in Fig. \ref{fig:AllSMassFrac} where we have plotted stellar mass fraction against total mass. One can clearly see the change in bias as one moves from older, colder galaxies with a high stellar mass fraction down to younger galaxies with lower stellar mass fractions. We only show galaxies with a stellar mass greater than $10^9 \, \mathrm{M_\odot}$ as the EAGLE model was not calibrated to reproduce the galaxy population below this threshold. The change in the bias at low masses between here and in Fig. \ref{fig:AllSMassSStel} is due to this stellar mass cut preferentially.

Binning galaxies by their stellar mass fraction does not reduce the scatter in the $\sigma-M$ relation compared to the stellar mass case as $N_\mathrm{eff,piv}=58$ and $77$ for the $[0.01,0.1]$ and $[0.1,1]$ stellar mass fraction bins respectively. These numbers are similar to the case for the stellar mass bins and so the scatter is also similar, with $\delta_{\ln}=0.11 \pm 0.01$ and $0.056 \pm 0.006$ respectively. We conclude that any reduction in the intrinsic scatter is masked by the limited number of galaxies per cluster.

In addition to binning galaxies by their stellar mass we also binned galaxies by their internal stellar velocity dispersion, $\varsigma_{\rm gal,*}$, and their maximum circular velocity, $V_{\rm c,Max}$. The fitting relations obtained when binning galaxies by $\varsigma_{\rm gal,*}$ and $V_{\rm c,Max}$ are given in Table~ \ref{z0ParamsManyRVelProx}. We find that binning by $\varsigma_{\rm gal,*}$ also returns an unbiased velocity dispersion. We conclude that selecting galaxies based on properties that are not significantly affected by entering the cluster environment is effective in mitigating the effects of velocity bias.

\subsection{Velocity bias and dynamical history} \label{FormHist}

We also need to explain the trend seen in Fig. \ref{fig:AllSMassSStel} between stellar mass fraction and velocity bias. To do this, we 
examine the dynamical history of the galaxies in more detail. We remind the reader that we define the time passed since a 
galaxy first entered inside $r_{\rm 200c}(z)$ of the main cluster as $t_{r_{\rm 200c}}$, so that a galaxy that entered at the present day has $t_{r_{\rm 200c}}=0$. 

Fig. \ref{fig:HM26vrt} shows the radial positions of all galaxies in CE-26 at $z=0$ that had a total mass 
$M_{\rm Sub}> 10^9 \, \mathrm{M_\odot}$ when they first entered $r_{\rm 200c}$ of the main halo, as a function of 
$t_{r_{\rm 200c}}$.\footnote{Fig. \ref{fig:HM26vrt} is qualitatively representative of what is observed across most clusters. 
Clusters that have recently undergone a major merger show more complex structure.} On the basis that these positions
are sampling typical orbital trajectories, we see that the vast majority of subhaloes will go beyond $r_{\rm 200c}$ after their first infall. 
However, when the galaxies fall back onto the cluster their motions are less coherent. (Note that the striping seen for $t_{r_{\rm 200c}}>7 \, \mathrm{Gyr}$ is due to the finite number of outputs from the simulation.) Recent work by \cite{Rhee2017} 
found similar structure when analysing subhalo positions in phase-space as a function of time. \cite{Rhee2017} found very weak 
dependence on galaxy and host cluster mass, and we find the same qualitative structure across the whole mass range of clusters studied in the C-EAGLE-GAS sample. 

\begin{figure}
  \centering
  \includegraphics[width=0.99\linewidth]{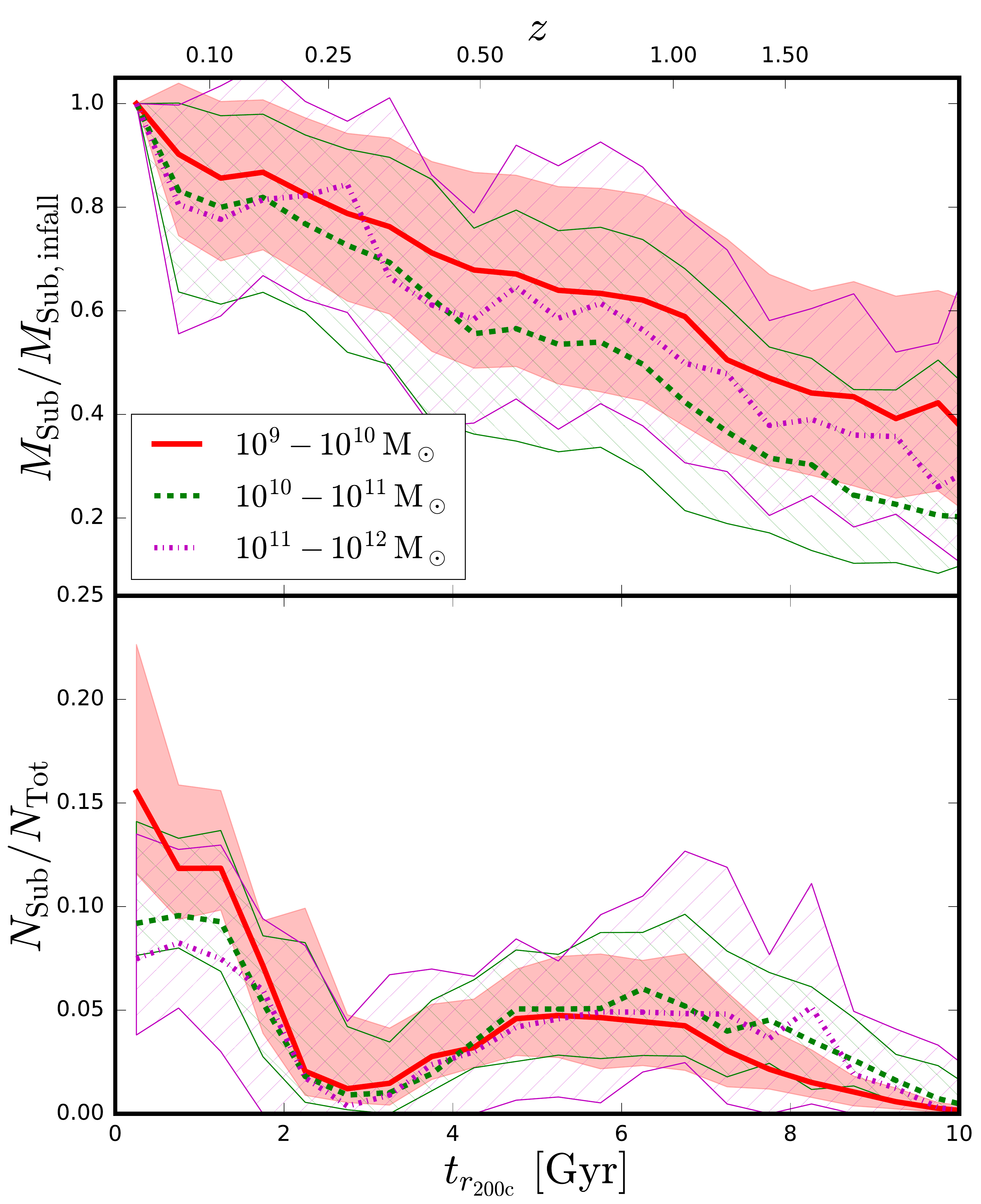}
  \caption{Top panel: ratio of a galaxy's present ($z=0$) mass to its mass when it first crossed $r_{\rm 200c}$. Bottom panel: 
  fraction of galaxies inside $r_{\rm 200c}$ at $z=0$ that fell into the cluster at $t_{r_{\rm 200c}}$. Results in both panels are 
  plotted against $t_{r_{\rm 200c}}$. The median values for all 30 clusters are shown as the bold lines with shaded regions depicting 
  the 16th and 84th percentiles. Galaxies are binned with respect to their total mass at infall.}
  \label{fig:MtotMinfallComp}
\end{figure}

The top panel of Fig. \ref{fig:MtotMinfallComp} shows the ratio of galaxy's total mass at $z=0$ to that at $t_{r_{\rm 200c}}$, versus 
$t_{r_{\rm 200c}}$. As expected, older galaxies have lost more mass. For example, after $\sim 10 \, \mathrm{Gyr}$ in a cluster, a 
galaxy will have lost $60-70$ per cent of its initial mass. The results are split across three total mass bins (measured at $t_{r_{\rm 200c}}$). There is little, if any, total mass dependence for the proportional rate of mass loss when a galaxy enters a cluster, although the results are biased towards those that have survived until the present day. We find the rate of mass loss to be broadly similar to that found by \cite{Joshi2017}.

The bottom panel of Fig. \ref{fig:MtotMinfallComp} shows the median fraction of galaxies as a function of $t_{r_{\rm 200c}}$, averaged across all 30 clusters. As expected from Fig. \ref{fig:HM26vrt}, there is a deficit of galaxies that first crossed $r_{\rm 200c}$ $\sim 3 \mathrm{Gyr}$ ago, as most of those galaxies are now beyond $r_{\rm 200c}$. We also see a slight total mass dependence, with a greater of proportion of the lower mass $10^9-10^{10} \, \mathrm{M_\odot}$ galaxies entering the cluster more recently than the higher mass $10^{11}-10^{12} \, \mathrm{M_\odot}$ galaxies. This is expected as lower-mass galaxies are closer to the resolution limit of the simulation and so are less likely to survive until the present day.

\begin{figure}
    \centering
    \includegraphics[width=0.99 \linewidth]{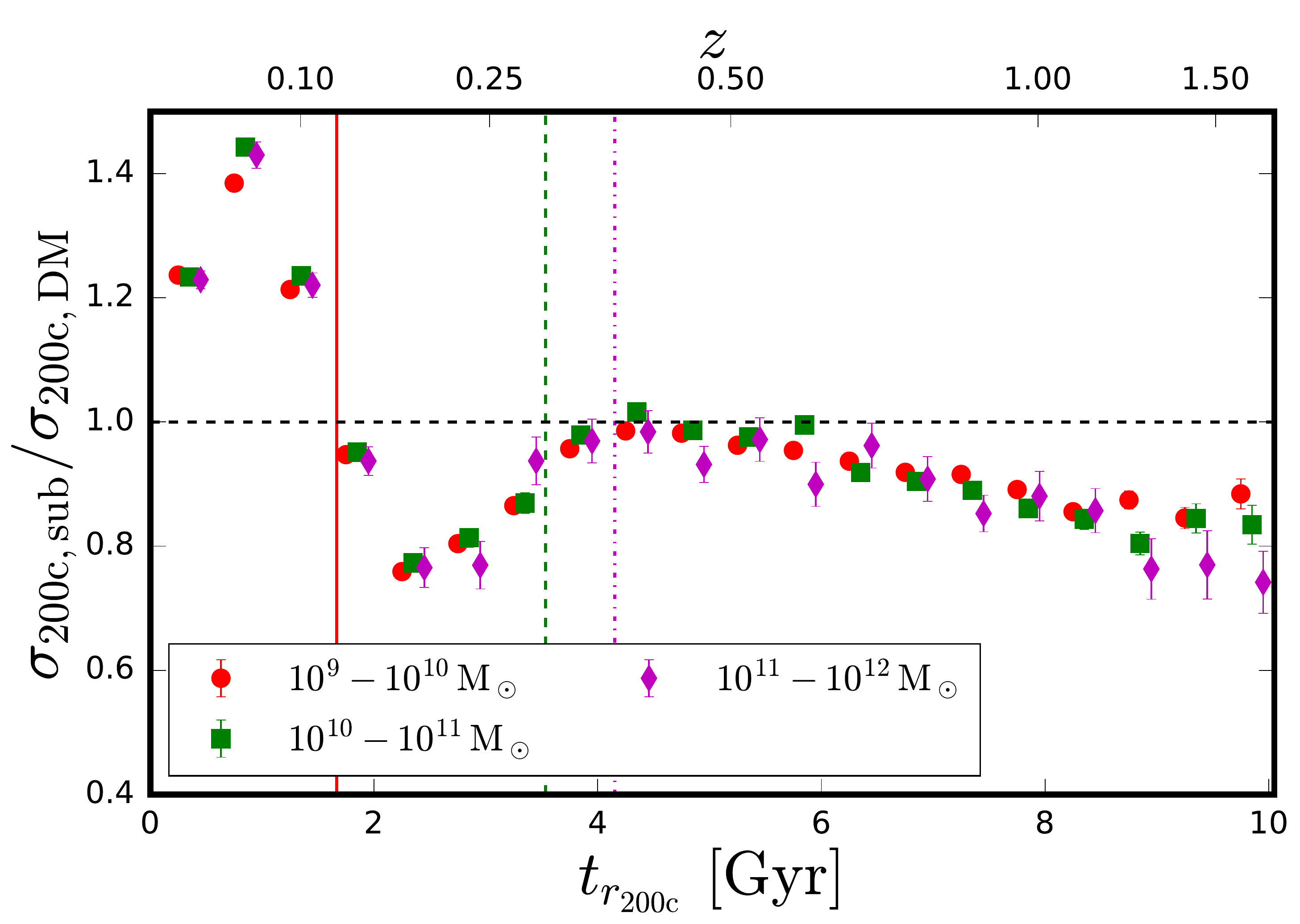}
    \caption{Velocity bias of galaxies inside $r_{\rm 200c}$ at $z=0$ shown as a function of time spent in the cluster. The velocities of the galaxies were divided by the velocity dispersion of the host cluster and all galaxies were then binned by their infall total mass. The error bars correspond to $\pm 1$ standard deviation, obtained by bootstrap re-sampling $10,000$ times. The vertical solid, dashed, and dot-dashed lines show the median value of $t_{r_{\rm 200c}}$ for the low, middle and high mass bins respectively.}
    \label{fig:sigmaTime}
\end{figure}

In Fig.~\ref{fig:sigmaTime} we show how the velocity bias, estimated from stacking all 30 clusters together, depends on infall time. Again, we do this by normalising the velocity of each galaxy to the velocity dispersion of the DM particles in its host halo; the galaxies are also split into three bins of varying total mass at infall. 

There are three key stages in this plot. Firstly, 
galaxies that entered the cluster in the last $2 \, \mathrm{Gyr}$ have a high velocity dispersion compared to the DM particles in 
their host cluster (i.e. $b_v>1$), with the peak bias coinciding with the point where they have had time to reach the pericentre of their first orbit
(see Fig.~\ref{fig:HM26vrt}). Secondly, galaxies that entered between $2-4 \, \mathrm{Gyr}$ ago are biased low ($b_v<1$); this 
corresponds to the minimum in the galaxy fraction, shown in the lower panel of Fig.~\ref{fig:MtotMinfallComp}. Most of these 
galaxies have gone beyond $r_{\rm 200c}$, with only the slower objects, or those on more circular orbits, remaining. Finally, at 
$4 \, \mathrm{Gyr}$, we see that $b_v \simeq 1$ and then a smooth trend towards a lower velocity bias with age, as the process 
of dynamical friction gradually slows the galaxies down.

Fig.~\ref{fig:sigmaTime} also implies that there is little total mass dependence in the slowdown of galaxies in the cluster when considering the 
last $\backsim 9 \, \mathrm{Gyr}$, contrary to what one might expect from analytical models \citep{Chandrasekhar1943,Adhikari2016}. 
Instead, the velocity bias originates from the distribution of galaxies as a function of $t_{r_{\rm 200c}}$. The vertical lines 
in Fig.~\ref{fig:sigmaTime} show the median values of $t_{r_{\rm 200c}}$ for the three mass bins, highlighting the total mass dependence seen in the lower panel of Fig.~\ref{fig:MtotMinfallComp}: lower mass galaxies tend to be younger than their higher 
mass counterparts. As the galaxies are only positively biased on their first pass through a cluster, the lower the 
$t_{r_{\rm 200c}}$, the higher the velocity bias. When $t_{r_{\rm 200c}}>9 \, \mathrm{Gyr}$, we do see signs of a total mass dependence in the velocity bias, with the velocity dispersion of the highest mass subhaloes decreasing faster than the others. The same qualitative shape shown in Fig. \ref{fig:sigmaTime} has also been observed in \cite{Ye2017} using the Illustris simulations.

\begin{figure}
    \centering
    \includegraphics[width=0.99 \linewidth]{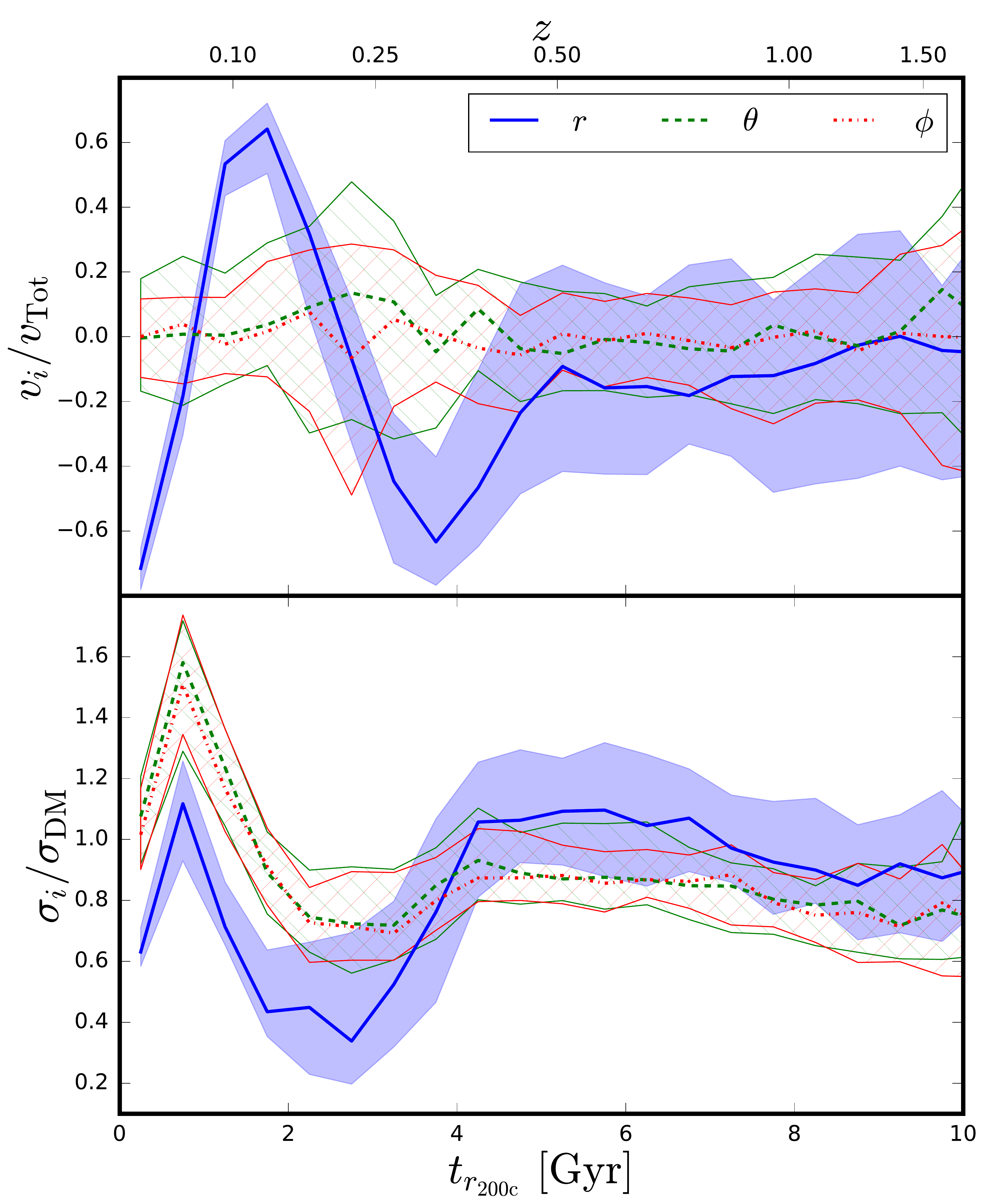}
    \caption{Radial and tangential velocity components (top panel) and velocity dispersion components (lower panel) for galaxies in all 30 clusters versus infall time, $t_{r_{\rm 200c}}$. Bold lines show the median values while the corresponding hatched regions represent the cluster-to-cluster scatter ($\pm 1$ standard deviation). The results are normalised to the velocity dispersion of the DM particles.}
    \label{fig:VSigBeta}
\end{figure}

We finally look at how the different components (radial and angular co-ordinates: $r, \theta, \phi$) 
of the galaxy's velocity depend on $t_{r_{\rm 200c}}$. 
Fig.~\ref{fig:VSigBeta} shows the median velocity component ($v_i$; top panel), velocity dispersion component 
($\sigma_i$; lower panel) as a function of $t_{r_{\rm 200c}}$ for all galaxies in the 30 clusters with $M_{\rm Sub}>10^9 \, \mathrm{M_\odot}$. The galaxy velocities are scaled by the velocity dispersion of the DM particles.

As expected, the tangential velocity components approximately average to zero and only the radial component has a strong 
directional preference over the last 4 Gyr. We can also see that there is very little spread between clusters in terms of their  
radial velocity before $\sim 4 \, \mathrm{Gyr}$, as the majority of galaxies are at the same point in their orbit. The velocity 
dispersion of each component, shown in the lower panel of Fig. \ref{fig:VSigBeta}, shows the same trend as Fig.~\ref{fig:sigmaTime}.
However, we see that the dispersion of the radial component is below the two tangential components 
for $t_{r_{\rm 200c}}<4 \, \mathrm{Gyr}$. Again, this is due to all of the galaxies accelerating towards the centre of the cluster 
(or being decelerated as they move outwards), reducing the width of the velocity distribution, whereas the the tangential 
components have no reason to be aligned by this effect.  

\subsection{Dependence on aperture size} \label{aperture_var}
\begin{figure}
    \centering
    \includegraphics[width=0.99 \linewidth]{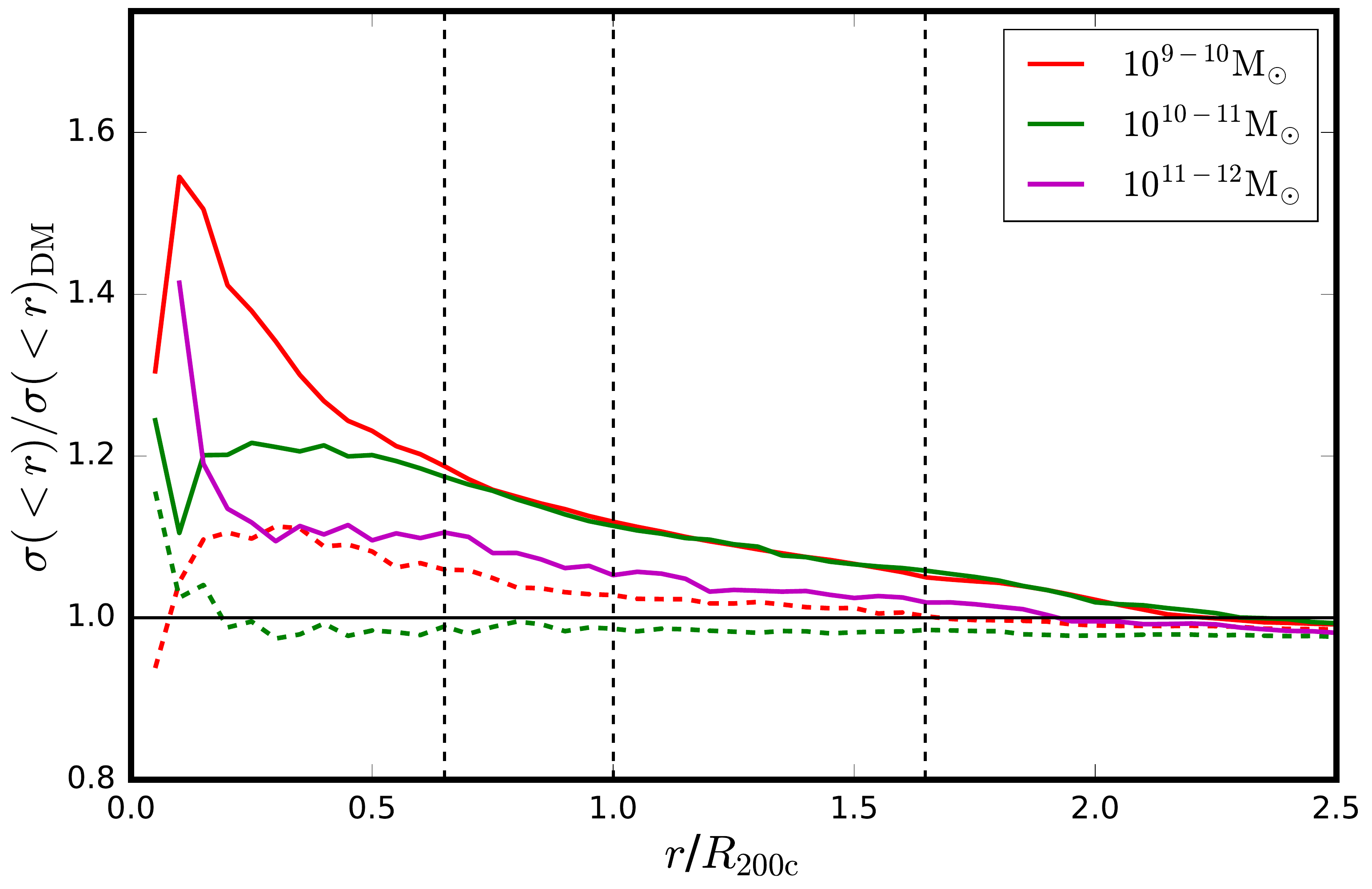}
    \caption{The cumulative velocity dispersion of the galaxies binned by total mass (solid lines) and stellar mass (dashed lines) as a function of distance from the cluster, relative to the cumulative velocity dispersion of the DM particles. The radial distance is in units of $r_{\rm 200c}$, with medium values of $r_{\rm 500c}$and $r_{\rm 200m}$ denoted by the vertical black dashed lines respectively as well as $r_{\rm 200c}$. The coloured lines show the median values for the 30 clusters.}
    \label{fig:CumSigma}
\end{figure}

Throughout this paper we have focussed on the velocity dispersion inside $r_{\rm 200c}$. We now present results for two other common apertures, $r_{\rm 500c}$ and $r_{\rm 200m}$, where the latter is defined relative to the mean density rather than the critical density. Fig. \ref{fig:CumSigma} shows the median cumulative $\sigma$ profiles, normalised to the cumulative velocity dispersion of the DM particles, using all galaxies within a given radius. We can see that the $\sigma(<r)/\sigma(<r)_{\mathrm{DM}}$ ratio changes significantly less when the galaxies are binned by stellar mass compared to total mass. We quantify this by calculating the $\sigma - M$ relation at $r_{\rm 500c}$ and $r_{\rm 200m}$, the parameters of which are given in Table. \ref{z0ParamsManyR}, where we have duplicated the relevant parameters for $r_{\rm 200c}$ for ease of comparison.

For the three apertures in Table \ref{z0ParamsManyR} we can see that $\sigma_{\mathrm{piv}}/\sigma_{\mathrm{piv,DM}}$ changes substantially more as a function of radius for the two lowest total mass bins than for the stellar mass bins. 
Beyond $2r_{\rm 200c}$ there is very little, if any bias, when selecting by either total or stellar mass, one would expect the bias to lessen with distance as processes such as dynamical friction will not be acting to the same degree as inside the cluster.
Qualitatively we see the same features in the $\sigma - M$ relation for the three apertures; selecting galaxies by total mass results in a positive velocity bias, where as the more observationally motivated selection by stellar mass yields little ($<5\%$) to no bias on average.

\begin{table}
    \centering
    \caption{Best-fit parameters for the $\sigma-M$ relation, for DM particles and galaxy total/stellar mass bins 
    within $r_{\rm 500c}$, $r_{\rm 200c}$ and  $r_{\rm 200m}$ at $z=0$. Symbols have the same meanings as 
    in Table~\ref{z0Params}.}
    \label{z0ParamsManyR}
    \begin{tabular}{lccccc}
         \hline
         $r_{\rm 500c}$ & $\alpha$ & $\sigma_{\mathrm{piv}} \, [\mathrm{km/s}]$ & $\sigma_{\mathrm{piv}}/\sigma_{\mathrm{piv,DM}}$  \\
         \hline
         DM Particles & $0.36 \pm 0.01$ &  $~958 \pm ~9 $ & \\ 
         $M_{\mathrm{Sub}}$ \\
         $10^9~-10^{10} \, \mathrm{M_\odot}$ & $0.33 \pm 0.01$ & $1101 \pm ~8 $ & $1.15 \pm 0.01$ \\ 
         $10^{10}-10^{11} \, \mathrm{M_\odot}$ & $0.33 \pm 0.01$ & $1093 \pm ~9 $ & $1.14 \pm 0.01$ \\ 
         $10^{11}-10^{12} \, \mathrm{M_\odot}$ & $0.33 \pm 0.02$ & $1037 \pm 13 $ & $1.08 \pm 0.02$\\ 
         $M_{*}$ \\
         $10^{9~}-10^{10} \, \mathrm{M_\odot}$ & $0.32 \pm 0.01$ & $~992 \pm ~9 $ & $1.04 \pm 0.01$\\ 
         $10^{10}-10^{11} \, \mathrm{M_\odot}$ & $0.36 \pm 0.02$ & $~941 \pm 12 $ & $0.98 \pm 0.02$\\
         \hline
         $r_{\rm 200c}$\\
         \hline
         DM Particles & $0.35 \pm 0.01$ &  $804 \pm ~9 $\\ 
         $M_{\mathrm{Sub}}$ \\
         $10^9~-10^{10} \, \mathrm{M_\odot}$ & $0.32 \pm 0.01$ & $883 \pm ~7 $ & $1.10 \pm 0.01$ \\ 
         $10^{10}-10^{11} \, \mathrm{M_\odot}$ & $0.33 \pm 0.01$ & $885 \pm ~8 $ & $1.10 \pm 0.02$ \\
         $10^{11}-10^{12} \, \mathrm{M_\odot}$ & $0.34 \pm 0.02$ & $844 \pm 11 $ & $1.05 \pm 0.02$ \\
         $M_{*}$ \,\,\, \\
         $10^{9~}-10^{10} \, \mathrm{M_\odot}$ & $0.32 \pm 0.02$ & $823 \pm 11 $ & $1.02 \pm 0.02$ \\
         $10^{10}-10^{11} \, \mathrm{M_\odot}$ & $0.36 \pm 0.02$ & $789 \pm 13 $ & $0.98 \pm 0.02$ \\
         \hline
         $r_{\rm 200m}$  \\
         \hline
         DM Particles & $0.35 \pm 0.01$ &  $683 \pm ~8 $\\ 
         $M_{\mathrm{Sub}}$ \\
         $10^9~-10^{10} \, \mathrm{M_\odot}$ & $0.335 \pm 0.009$ & $710 \pm ~6 $ & $1.04 \pm 0.02$ \\ 
         $10^{10}-10^{11} \, \mathrm{M_\odot}$ & $0.344 \pm 0.009$ & $710 \pm ~6 $ & $1.04 \pm 0.02$  \\ 
         $10^{11}-10^{12} \, \mathrm{M_\odot}$ & $0.34 \pm 0.01$ & $690 \pm 10 $ & $1.01 \pm 0.02$\\ 
         $M_{*}$ \,\,\, \\
         $10^{9~}-10^{10} \, \mathrm{M_\odot}$ & $0.33 \pm 0.02$ & $684 \pm 11 $ & $1.00 \pm 0.02$ \\ 
         $10^{10}-10^{11} \, \mathrm{M_\odot}$ & $0.35 \pm 0.02$ & $663 \pm 14 $ & $0.97 \pm 0.02$ \\
         \hline
    \end{tabular}
\end{table}

\section{Discussion and Conclusions} \label{discuss}

We have used the C-EAGLE simulations \citep{Bahe2017,Barnes2017b} to explore the extent and nature of the velocity bias introduced when using galaxies 
as dynamical tracers of the underlying cluster potential. The simulations are amongst the highest resolution 
cosmological simulations of galaxy clusters run to date (1 kpc force resolution at $z \approx 0$) and adopt a subgrid physics prescription that has 
been calibrated to produce the stellar mass functions, sizes, and black hole masses of field galaxies \citep{Schaye2015,Crain2015}. Recent work has also demonstrated that 
the C-EAGLE simulations approximately reproduce key properties of the galaxies \citep{Bahe2017} and hot gas \citep{Barnes2017b} in cluster environments.

In line with other recent work, we have found that DM-only subhaloes have a velocity dispersion that is around 5-10 per cent higher than 
the DM particles within $r_{\rm 200c}$. However, this bias is significantly reduced when selecting galaxies in hydrodynamical simulations based on their 
observational properties such as stellar mass or stellar velocity dispersion. This finding has only been possible because of the 
high resolution of the C-EAGLE galaxies, where the stellar component, located at the subhalo centre of potential, is more robust to 
stripping effects than the surrounding DM component. This may explain the difference between our results and those of other groups such a \cite{Munari2013}, who found a similar bias when binning in total and stellar mass. 

Our main findings can be summarised as follows:
\vspace{-0.1cm}
\begin{enumerate}
\item The intrinsic velocity dispersion of the DM particles in a cluster is tightly correlated with cluster masss, with and without 
the inclusion of baryonic physics (Fig.~\ref{fig:SigmavsmassDMPonly029}). The relation obtained in this paper is consistent with 
previous work by \cite{Evrard2008} and \cite{Munari2013}. 

\item In the C-EAGLE-DMO simulations, the velocity dispersion of subhaloes has a relatively constant bias, 
$\sim 10$ per cent above the DM particle value, regardless of subhalo mass 
($10^{9}<M_{\rm Sub}/\mathrm{M_{\odot}}<10^{12}$) or redshift ($0 \leq z \leq 1.5$).
The inclusion of baryonic physics has little effect on the velocity dispersion of low mass galaxies but higher mass 
($>10^{11} \, \mathrm{M_\odot}$) galaxies have their bias reduced to around 5 per cent 
(Fig.~\ref{fig:ParameterComp}).

\item The velocity bias is suppressed to within a few per cent when selecting galaxies based on their stellar mass instead of their 
total mass (Fig.~\ref{fig:ParameterComp}). This difference arises from having a larger fraction of galaxies that have newly entered the 
cluster when selecting by total mass; these galaxies tend to be hotter than objects that have been in the cluster for longer (Figs. \ref{fig:AllSMassSStel} and \ref{fig:AllSMassFrac}). 

\item We find that the fractional mass loss of galaxies has an approximately linear trend with the time spent in the cluster, 
with little dependence on the cluster or total mass, in agreement with \cite{Joshi2017} and \cite{Rhee2017}. We also find that a 
marginally larger fraction of low mass galaxies have newly entered the clusters relative to high mass galaxies. 
This would be expected partly due to the limitations of SUBFIND not being able to reliably identify galaxies 
$\lesssim 10^8 \,  \mathrm{M_\odot}$ (Fig. \ref{fig:MtotMinfallComp}) \citep{Muldrew2011}.

\item Figs. \ref{fig:HM26vrt} and \ref{fig:MtotMinfallComp} can be used together to explain why velocity bias has a strong dependence on infall time,  $t_{r_{\rm 200c}}$, as shown in Fig. \ref{fig:sigmaTime}. Galaxies on their first pass through a cluster are on highly 
radial orbits, with the peak in velocity dispersion occurring closest to the pericentre, i.e. the cluster centre. Once the galaxies enter $r_{\rm 200c}$ for the second time they are approximately unbiased and their velocity dispersion steadily decreases due to 
dynamical friction thereafter. 

\item The velocity bias varies more strongly as a function distance from the centre of the cluster when binning galaxies by total mass compared to stellar mass. At $r_{\rm 500c}$, $r_{\rm 200c}$ and $r_{\rm 200m}$ binning galaxies by stellar mass results in a bias of less than 5 per cent (Fig. \ref{fig:CumSigma}).

\end{enumerate}

In conclusion, we find that the velocity bias of cluster galaxies within $r_{\rm 200c}$ is small, $b_v =1 \pm 0.05$, 
out to beyond $z=1$, so long as the galaxies are selected by their (observable) stellar mass or velocity dispersion. This has promising 
implications for cluster cosmology where a growing number of new spectroscopic surveys will be used to measure 
cluster masses via galaxy kinematics, e.g. using the caustic method \citep{Diaferio1997,Gifford2017}. 

In future work, 
we plan to use the C-EAGLE cluster sample to investigate the reliability of caustic mass estimates. We will also 
simulate observations of these clusters to establish how observational considerations such as line of sight contamination, 
selection bias and projection effects affect the results, as well as how more observationally expensive techniques can be 
combined with the $\sigma - M$ relation to yield tighter mass constraints. 

\section*{Acknowledgements}
We would like to thank Ian McCarthy and Joop Schaye for their helpful comments on the initial manuscript.
This work used the DiRAC Data Centric system at Durham University, operated by
the Institute for Computational Cosmology on behalf of the STFC DiRAC HPC
Facility (www.dirac.ac.uk). This equipment was funded by BIS National
E-infrastructure capital grant ST/K00042X/1, STFC capital grants ST/H008519/1
and ST/K00087X/1, STFC DiRAC Operations grant ST/K003267/1 and Durham
University. DiRAC is part of the National E-Infrastructure.
The Hydrangea simulations were in part performed on the German federal maximum performance computer 
"HazelHen" at the maximum performance computing centre Stuttgart (HLRS), under project GCS-HYDA / ID 44067 financed through the large-scale project "Hydrangea" of the Gauss Center for Supercomputing. Further simulations were performed at the Max Planck Computing and Data Facility in Garching, Germany. 
We also gratefully acknowledge PRACE for awarding the EAGLE project access to the Curie facility based in France at Tr\'es Grand Centre de Calcul. DJB and STK acknowledge support from STFC through grant ST/L000768/1. TJA is supported by an STFC studentship.



\bibliographystyle{mnras}
\bibliography{main} 



\appendix

\section{Additional cluster properties}

\begin{table}
    \centering
    \caption{Global properties of the C-EAGLE clusters at $z=0$, used in this paper. The velocity dispersion 
    $\sigma_{\rm 200c}$ is calculated using only the DM particles inside $r_{\rm 200c}$. $r_\mathrm{clean}$ 
    is the maximum radius containing only high resolution particles.}
    \label{Table:ClusterProps}
    \begin{tabularx}{\columnwidth}{l c c c c}
        \hline
        Halo & $r_{\rm 200c} \, [\mathrm{Mpc}]$ & $M_{\rm 200c} \, [\mathrm{M_\odot}]$ & $\sigma_{\rm 200c} \, [\mathrm{km/s}]$ & $r_\mathrm{clean}/r_{\rm 200c}$ \\
        \hline
        CE-00 & 1.04 & $1.19 \times 10^{14}$ & 457 & 5 \\
        CE-01 & 1.02 & $1.15 \times 10^{14}$ & 425 & 5 \\
        CE-02 & 1.05 & $1.22 \times 10^{14}$ & 462 & 5 \\
        CE-03 & 1.08 & $1.36 \times 10^{14}$ & 476 & 5 \\
        CE-04 & 1.19 & $1.78 \times 10^{14}$ & 544 & 5 \\
        CE-05 & 1.09 & $1.38 \times 10^{14}$ & 517 & 5 \\
        CE-06 & 1.27 & $2.20 \times 10^{14}$ & 548 & 10 \\
        CE-07 & 1.27 & $2.17 \times 10^{14}$ & 553 & 10 \\
        CE-08 & 1.30 & $2.37 \times 10^{14}$ & 584 & 5 \\
        CE-09 & 1.39 & $2.86 \times 10^{14}$ & 633 & 5 \\
        CE-10 & 1.40 & $2.94 \times 10^{14}$ & 617 & 5 \\
        CE-11 & 1.44 & $3.19 \times 10^{14}$ & 682 & 5 \\
        CE-12 & 1.55 & $3.96 \times 10^{14}$ & 683 & 10 \\
        CE-13 & 1.57 & $4.11 \times 10^{14}$ & 768 & 10 \\
        CE-14 & 1.62 & $4.55 \times 10^{14}$ & 672 & 10 \\
        CE-15 & 1.71 & $5.31 \times 10^{14}$ & 737 & 10 \\
        CE-16 & 1.74 & $5.62 \times 10^{14}$ & 880 & 10 \\
        CE-17 & 1.65 & $4.78 \times 10^{14}$ & 682 & 5 \\
        CE-18 & 1.87 & $6.94 \times 10^{14}$ & 849 & 10 \\
        CE-19 & 1.86 & $6.84 \times 10^{14}$ & 868 & 5 \\
        CE-20 & 1.77 & $5.96 \times 10^{14}$ & 823 & 5 \\
        CE-21 & 2.00 & $8.55 \times 10^{14}$ & 873 & 5 \\
        CE-22 & 2.14 & $1.04 \times 10^{14}$ & 938 & 10 \\
        CE-23 & 1.99 & $8.39 \times 10^{14}$ & 936 & 5 \\
        CE-24 & 2.27 & $1.24 \times 10^{15}$ & 1107 & 10 \\
        CE-25 & 2.36 & $1.40 \times 10^{15}$ & 1005 & 10 \\
        CE-26 & 2.39 & $1.45 \times 10^{15}$ & 1119 & 5 \\
        CE-27 & 2.39 & $1.46 \times 10^{15}$ & 1264 & 5 \\
        CE-28 & 2.50 & $1.67 \times 10^{15}$ & 1178 & 10 \\
        CE-29 & 2.82 & $2.39 \times 10^{15}$ & 1223 & 10 \\
        \hline
    \end{tabularx}
\end{table}

In this Appendix, we list additional cluster properties that the reader may find useful. In Table~\ref{Table:ClusterProps} we list the measured velocity dispersions of the DM particles within $r_{\rm 200c}$ for each C-EAGLE cluster at $z=0$. These are presented alongside $r_{\rm 200c}$, $M_{\rm 200c}$ and the size of the high-resolution region. The Table can be used in conjunction with additional data presented in \cite{Bahe2017} and \cite{Barnes2017b}.

\begin{table}
    \centering
    \caption{Best-fit parameters to the $\sigma_{\rm 200c}-M_{\rm 200c}$ relation for the DM particles and galaxies at $z=0$ with the logarithmic slope, $\alpha$, fixed to the self-similar value of $1/3$. Column~1 lists the simulation and sample details (with mass limits where appropriate). 
    Columns 2-3 give the 
    best-fit normalisation and scatter. Finally, column 4 gives the ratio of the 
    normalisation to the case for DM particles, a measure of the velocity bias for the galaxies.}
    \label{z0Params_alphaFixed}
    \begin{tabular}{l c c c c}
        \hline
         & $\sigma_{\mathrm{piv}} \, [\mathrm{km/s}] $ & $\delta_{\ln}$ & $\sigma_{\mathrm{piv}}/\sigma_{\mathrm{piv,DM}}$\\ 

         \hline
    
    C-EAGLE-DMO\\
         DM Particles & $803 \pm 7\phantom{0}$ & $0.048  \pm  0.006$\\ 

         $M_\mathrm{Sub}:$ $10^9~-10^{10} \,  \mathrm{M_\odot}$  &  $882 \pm 6\phantom{0}$ & $0.036  \pm  0.005$ & $1.10\pm0.01$\\ 

         $M_\mathrm{Sub}:$ $10^{10}-10^{11} \, \mathrm{M_\odot}$ &  $891 \pm 6\phantom{0}$  & $0.035  \pm  0.005$ & $1.11\pm0.01$\\ 

         $M_\mathrm{Sub}:$ $10^{11}-10^{12} \, \mathrm{M_\odot}$ &  $891 \pm 10$  & $0.061  \pm  0.008$ & $1.11\pm0.02$\\

         \hline

    C-EAGLE-GAS\\
         DM Particles & $800\pm9\phantom{0}$ & $0.058  \pm  0.007$\\ 

         $M_\mathrm{Sub}:$ $10^9~-10^{10} \, \mathrm{M_\odot}$ & $880\pm6\phantom{0}$ & $0.034  \pm  0.005$ & $1.10\pm0.01$\\ 

         $M_\mathrm{Sub}:$ $10^{10}-10^{11} \, \mathrm{M_\odot}$ & $883\pm6\phantom{0}$ & $0.035  \pm  0.004$ & $1.10\pm0.02$\\ 

         $M_\mathrm{Sub}:$ $10^{11}-10^{12} \, \mathrm{M_\odot}$ & $847\pm10$ & $0.09\phantom{0}  \pm  0.02\phantom{0}$ & $1.06\pm0.02$\\ 

         $M_{\mathrm{*}}\,\,\,\,\,:$ $10^{9~}-10^{10} \, \mathrm{M_\odot}$ & $820\pm10$ & $0.062  \pm  0.007$ & $1.02\pm0.02$\\ 

         $M_{\mathrm{*}}\,\,\,\,\,:$ $10^{10}-10^{11} \, \mathrm{M_\odot}$ & $796\pm11$ & $0.11\phantom{0}  \pm  0.02\phantom{0}$ & $1.00\pm0.02$\\

         \hline

    \end{tabular}
\end{table}

\begin{table}
    \label{Table:MnyRValsVelProx}
    \centering
    \caption{Best-fit $\sigma-M$ parameters for the subhaloes when binned by 
    stellar velocity dispersion and maximum circular velocity.}
    \label{z0ParamsManyRVelProx}
    \begin{tabularx}{\columnwidth}{lccccc}
         \hline
         $r_{\rm 200c}$ & $\alpha$ & $\sigma_{\mathrm{piv}} \, [\mathrm{km/s}]$ & $\sigma_{\mathrm{piv}}/\sigma_{\mathrm{piv,DM}}$  \\
         \hline
         $\varsigma_\mathrm{gal,*}$\,\,\,\,\,: \,\,\,$50 - 100 \mathrm{km/s}$ & $0.33 \pm 0.02$ & $825 \pm 14 $ & $1.03 \pm 0.02$ \\
         $\varsigma_\mathrm{gal,*}$\,\,\,\,\,: $100 - 150 \mathrm{km/s}$ & $0.36 \pm 0.03$ & $779 \pm 18 $ & $0.97 \pm 0.03$ \\
                  $\varsigma_\mathrm{gal,*}$\,\,\,\,\,: $150 - 200 \mathrm{km/s}$ & $0.36 \pm 0.05$ & $828 \pm 32 $ & $1.03 \pm 0.04$ \\
         \hline
         $V_\mathrm{c,Max}$ : \,\,\,$50 - 100 \mathrm{km/s}$ & $0.33 \pm 0.01$ & $876 \pm ~9 $ & $1.09 \pm 0.02$ \\
         $V_\mathrm{c,Max}$ : $100 - 150 \mathrm{km/s}$ & $0.34 \pm 0.02$ & $835 \pm 14 $ & $1.04 \pm 0.02$ \\
         $V_\mathrm{c,Max}$ : $150 - 200 \mathrm{km/s}$ & $0.32 \pm 0.03$ & $814 \pm 18 $ & $1.02 \pm 0.03$ \\
         $V_\mathrm{c,Max}$ : $200 - 250 \mathrm{km/s}$ & $0.40 \pm 0.03$ & $750 \pm 20 $ & $0.94 \pm 0.03$ \\
         \hline
    \end{tabularx}
\end{table}

Table~\ref{z0Params_alphaFixed} shows the fitting parameters as in Table~\ref{z0Params} but with $\alpha$ fixed to the self-similar value of $1/3$. We find no significant change in the normalisation, scatter or bias when the gradient is fixed.

In Table~\ref{z0ParamsManyRVelProx}, we show the best-fit parameters when binning galaxies by their stellar 
velocity dispersion $\varsigma_\mathrm{gal,*}$ as well as their maximum circular velocity, $V_\mathrm{c,Max}$, within $r_{\rm 200c}$. In general, the results show a similarly small bias as with the stellar mass bins.


\bsp    
\label{lastpage}
\end{document}